\documentclass[manuscript, review=false, screen=true]{acmart}

\AtBeginDocument{%
  }

\setcopyright{acmcopyright}
\copyrightyear{2023}
\acmYear{2023}
\acmDOI{XXXXXXX.XXXXXXX}

\acmJournal{JACM}
\acmVolume{0}
\acmNumber{0}
\acmArticle{000}
\acmMonth{00}

\usepackage{algorithm}
\usepackage{algpseudocode}
\usepackage{wrapfig}

\usepackage{enumitem}
\setitemize{noitemsep,topsep=1pt,parsep=0pt,partopsep=1pt}

\makeatletter
\algrenewcommand\ALG@beginalgorithmic{\footnotesize}
\makeatother

\usepackage[normalem]{ulem}

\newcommand{\remark}[1]{\textcolor{yellow}{#1}}
\newcommand{\removed}[1]{\leavevmode{\color{red}{\sout{#1}}}}

\def \cleanversion{} %
\ifx\cleanversion\undefined
\else
 \renewcommand{\remark}[1]{} %
 \renewcommand{\removed}[1]{} 
 
\fi

\usepackage{calc}
\newlength\myheight
\newlength\mydepth
\settototalheight\myheight{Xygp}
\newcommand*\inlinegraphics[1]{%
  \settototalheight\myheight{Xygp}%
  \settodepth\mydepth{Xygp}%
  \raisebox{-\mydepth}{\includegraphics[height=\myheight]{#1}}%
}

\begin{document}

\title{A Spatial Constraint Model for Manipulating Static Visualizations}

\author{Can Liu}
\email{can.liu@pku.edu.cn}
\affiliation{%
  \streetaddress{Key Laboratory of Machine Perception (Ministry of Education), School of Intelligence Science and Technology}
  \institution{Peking University}
  \city{Beijing}
  \country{China}
  \postcode{100871}
}

\author{Yu Zhang}
\affiliation{%
  \institution{University of Oxford}
  \city{Oxford}
  \country{UK}}
\email{yu.zhang@cs.ox.ac.uk}

\author{Cong Wu}
\affiliation{%
\streetaddress{Key Laboratory of Machine Perception (Ministry of Education), School of Intelligence Science and Technology}
  \institution{Peking University}
  \city{Beijing}
  \country{China}
}

\author{Chen Li}
\affiliation{%
 \institution{Central Academy of Fine Arts}
 \city{Beijing}
 \country{China}}
 \email{chen.li.cafa@gmail.com}

\author{Xiaoru Yuan}
\affiliation{%
\streetaddress{Corresponding Author, Key Laboratory of Machine Perception (Ministry of Education), School of Intelligence Science and Technology, National Engineering Laboratory for Big Data Analysis and Application}
  \institution{Peking University}
  \state{Beijing}
  \country{China}}

\renewcommand{\shortauthors}{Liu et al.}

\begin{abstract}
  We introduce a spatial constraint model to characterize the positioning and interactions in visualizations, thereby facilitating the activation of static visualizations.
  Our model provides users with the capability to manipulate visualizations through operations such as selection, filtering, navigation, arrangement, and aggregation.
  Building upon this conceptual framework, we propose a prototype system designed to activate pre-existing visualizations by imbuing them with intelligent interactions. This augmentation is accomplished through the integration of visual objects with forces. The instantiation of our spatial constraint model enables seamless animated transitions between distinct visualization layouts. To demonstrate the efficacy of our approach, we present usage scenarios that involve the activation of visualizations within real-world contexts.
\end{abstract}

\begin{CCSXML}
    <ccs2012>
    <concept>
    <concept_id>10003120.10003123</concept_id>
    <concept_desc>Human-centered computing~Interaction design</concept_desc>
    <concept_significance>500</concept_significance>
    </concept>
    <concept>
    <concept_id>10003120.10003145.10003146</concept_id>
    <concept_desc>Human-centered computing~Visualization techniques</concept_desc>
    <concept_significance>300</concept_significance>
    </concept>
    </ccs2012>
\end{CCSXML}
    
\ccsdesc[500]{Human-centered computing~Interaction design}
\ccsdesc[300]{Human-centered computing~Visualization techniques}

\keywords{intelligent interaction, constraint}


 \maketitle

 \section{Introduction}

Interaction plays a crucial role in facilitating a dialogue between users and visualizations, providing alternative representations and different perspectives during the process of data exploration. Despite its importance, many visualizations in the real world are static or have limited interactive capabilities. This issue arises from both the creators' and users' perspectives. On the creators' side, traditional paper-based practices and the specialized knowledge required for visualization programming (using tools like $D^3$~\cite{bostock2011d3}, Vega~\cite{satyanarayan2015reactive}, or ECharts~\cite{li2018echarts}) contribute to the lack of interactivity in visualizations. On the users' side, complex interaction requirements beyond predefined options can also hinder the realization of interactive visualizations.

A comprehensive interaction process compasses both user actions and visualization changes.
Spatial change is one of the most prevalent types of changes, given that the spatial channel is one of the most perceptually effective visual channels~\cite{mackinlay1986APT}.
Therefore, many interactions can be categorized as spatial changes affecting elements within the visualization.
For example, re-encoding a grouped bar chart as a stacked bar chart or changing the stacking order of a ThemeRiver~\cite{havre2002themeriver} both fall under spatial transformations within the visualization, altering the position, shape, or size of visual elements.
Considering the prevalence of static visualizations and their potential benefits from interaction, we propose a spatial constraint model that can facilitate spatial interactions within visualizations.
This model involves defining spatial constraints on visual objects within the visualization, enabling user interactions with the original static visualizations, and supporting a wide range of interactions with them.

In the spatial constraint model, we conceptualize visual objects (such as points, bars, areas, etc.) as entities governed by physical principles. These visual objects are subjected to a set of spatial constraints, which encompass gravity, support, collision, and fixation constraints. These constraints are integrated to facilitate the construction of intricate interactions, forming the foundation for the sequence of visual object updates prompted by interactions. In contrast to conventional tool kits (e.g., D3~\cite{bostock2011d3}) that treat the positioning and shaping of visual objects as outcomes of pre-programmed functions, we adopt a physical perspective to model the spatial channels of visual objects. Crucially, our approach introduces spatial updates that are activated by interactions. From this standpoint, the stable positioning of visual objects emerges from the equilibrium achieved through the interplay of spatial constraints, drawing inspiration from the concept of object equilibrium.

To validate the spatial constraint model, we have developed a prototype system designed for the interactive manipulation of existing visualizations.
Following manipulation, the visualizations are transformed to converge towards new layouts, enhancing users ability to explore the visualization and underlying data.
These interactions facilitate a range of user tasks, such as reordering visual objects through dragging to alter stacking orders or orientations, zooming or dragging scales to reorganize or rescale axes, introducing new constraints, or modifying existing ones.
These operations enable intelligent interactions to support user analysis tasks including navigation, filtering, reordering, re-encoding, and aggregation. Subsequent to user interactions with the visualization, our system maps spatial constraints to forces and propels visual objects toward convergent positions using force-directed optimization. The resultant convergence leads to the emergence of a new visualization layout. Our model is applicable to various common visualization types, such as bar charts, area charts, bubble charts, and line charts.
We demonstrate the efficacy of our approach in supporting diverse interactive tasks for common visualizations through real-world examples and user studies.
We conducted user experiments with participants who have varying levels of expertise in visualization. The experiments demonstrated that the interactions supported by the spatial constraint model are intuitive and easy to understand, and they can facilitate diverse forms of interaction. Therefore, our approach is beneficial for both novice and expert users.
The contributions of this work encompass the following:
\begin{itemize}
\item A spatial constraint model that effectively represents both the positioning and positional changes of visual objects, while also facilitating intuitive transitions after user manipulations of visualizations. 
\item A prototype system grounded in the constraint model, enabling the activation and manipulation of static visualizations by users.
\end{itemize}

 \section{Related Work}
\label{section:background}

We contribute a spatial constraint approach to model visual objects, which is related to the direct manipulation of visual elements, interaction modeling in visualization, and the concept of force-directed layouts.

\subsection{Direct Manipulation for Visual Objects}

Direct manipulation provides an interface with continuous representation, physical actions, and ongoing feedback~\cite{directManipulation}, and it is more intuitive than traditional interactions, as it leverages physical concepts in the interface design.
Some approaches incorporate force metaphors in visualization; for example, the magnet metaphor~\cite{soo2005dust} and attractive and repulsive forces~\cite{rzeszotarski2014kinetica} are utilized to assist users in interacting with multivariate data items.
However, these approaches primarily focus on unit visualization; they treat each point as an individual data item.
Other studies use force metaphors to support visualization tasks.
For instance, Tominski et al.~\cite{tominski2012interaction} employed a force metaphor to facilitate folding interactions for comparisons.
While many prior approaches have integrated physical forces into visualization, where each data item is represented as a point (e.g., in graph layouts or unit visualizations), our approach is more generic; it applies constraints to various types of visual objects (e.g., bars and areas).
Saket et al.~\cite{saket2019investigating} proposed a space for manipulation of existing visualizations, such as resizing, recoloring, and repositioning. 
In Saket et al.'s approach, some operations like recoloring or repositioning actually introduce user-annotated data, which might lead to some changes in the data.
However, The direct manipulations in our model emphasize not altering the underlying data.

\subsection{Interaction Modeling for Visualization}

Visualization authoring tools, such as D3~\cite{bostock2011d3}, Vega-Lite~\cite{satyanarayan2016vegalite}, and ECharts~\cite{li2018echarts}, offer interaction features that enable users to initiate updates of visual objects in response to actions.
Nebula~\cite{chen2021nebula} introduces a grammar for modeling multi-view interactions.
Some other methods~\cite{choi2015visdock, harper2014deconstructing} aim to enhance existing visualizations through interactions.
VisDock~\cite{choi2015visdock} permits users to incorporate interactions (e.g., selection, filtering, navigation, etc.) into established visualizations using code.
Harper and Agrawala~\cite{harper2014deconstructing}, leveraging the characteristics of D3's DOM element specifications, deconstruct existing D3 visualizations by aligning provided data with visual attributes.
Moreover, Harper and Agrawala~\cite{harper2018converting} extract D3 visualizations and transform them into templates for reusability.
These methods demand users to be familiar with the authoring tool and coding of an existing visualization.
In contrast, our approach treats visual objects as physical entities with spatial constraints, making it agnostic to the specific authoring tool kits.

The utilization of physical force metaphors in visualization has been a subject of exploration over the past decades. Huron et al.~\cite{Huron2013Visual} introduced a sedimentation metaphor for visualization.
Saket et al.~\cite{saket2016visualization} reconfigure visualization layouts by directly manipulating visual objects and deducing the mapping of data attributes.
The approach proposed in this method can accommodate pre-existing visualizations and render them interactive, thus enhancing their utility.
In contrast, other methods are tailored to visualizations developed within their specific systems.
The methodology presented in this manuscript introduces a novel interpretation for the positioning of visual objects, their motion dynamics, and the mechanics of visual element movement in pre-existing visualizations.
This manuscript puts forward a scheme to guide the orchestrated motion of these visual elements after manipulation. Unlike other approaches that rely on metaphorical constructs for visualization design, our approach distinguishes itself by orchestrating the autonomous relocation of visual objects to strategically determined positions through the application of forces. Remarkably, these positions inherently encapsulate an effective mode of visual representation. From this perspective, the approach advocated in this study exhibits heightened generality and adaptability.

Interaction+~\cite{lu2017interaction} is also tool-agnostic; it parses the attributes of visual objects and applies additional interaction components to the visualization.
However, Interaction+ primarily targets non-spatial attributes such as color and opacity, which cannot be utilized to update the spatial positions of visual objects.

\subsection{Force-Directed Layout}

Force-directed algorithms have found extensive application in graph layout visualizations~\cite{forcedirected}.
The spring-electrical model, which combines attractive and repulsive forces, was initially introduced by Eades~\cite{eades1984heuristic}.
Building upon Eades's work~\cite{eades1984heuristic}, Fruchterman and Reingold~\cite{fruchterman1991graph} proposed the spring-embedder model.
Furthermore, Kamada and Kawai~\cite{kamada1989algorithm} treated the force-directed layout problem as an optimization challenge.
The force-directed graph layout problem can be reformulated into a constrained energy minimization problem~\cite{dwyer2005dig, gansner2004graph, zheng2018graph}.
Certain approaches~\cite{hu2005efficient, hachul2004drawing} introduce multi-level techniques to reduce computational complexity; these methods yield efficient and high-quality outcomes for drawing large graphs.
Moreover, users might impose customized constraints on graph nodes, such as fixing a node's position or maintaining a constant distance between two nodes.
Numerous approaches~\cite{kamps1995constraint, dwyer2006ipsep} have been developed to address such tailored constraints.

Apart from calculating graph layouts, the concept of constraints has been widely employed for interactive interaction; for instance, it can be harnessed to create a graph interface~\cite{garnet} and specify interactive objects~\cite{carr1994specification}.
In our approach, spatial constraints are utilized to guide visual objects toward equilibrium states.

\section{Method Overview}
\label{section:overview}

User interactions trigger changes in visual representations.
As per Mackinlay~\cite{mackinlay1986APT}, spatial channels (position, size, shape) are deemed the most effective channels.
Consequently, many changes prompted by interaction can be conceptualized as spatial modifications of visual objects.
We can employ Brehmer and Munzner's~\cite{Brehmer2013TaskAbstract} classification of interaction to explore how numerous interactions can be interpreted as spatial changes.
According to Brehmer and Munzner, interaction techniques applied to existing visual objects fall under the category of manipulation, encompassing actions like selecting, navigating, arranging, changing, filtering, and aggregating visual objects.

\begin{itemize}
\item \textbf{Rearrange current visual objects:} A substantial portion of visualization interactions revolves around adjusting the spatial layout of existing visual elements. For example, resizing a logarithmic axis to facilitate navigation within a visualization or transforming a stacked bar chart into a grouped bar chart for altered visual representation. Sorting elements within a visualization also falls into this category.
\item \textbf{Reduce current visual objects:} Certain interactions aim to diminish information content. For instance, actions such as selecting and filtering. Selection doesn't modify the visualization's content, whereas filtering or removing a subset of visual objects does.
\end{itemize}

\begin{figure*}[!ht]
    \centering
    \includegraphics[width=\textwidth]{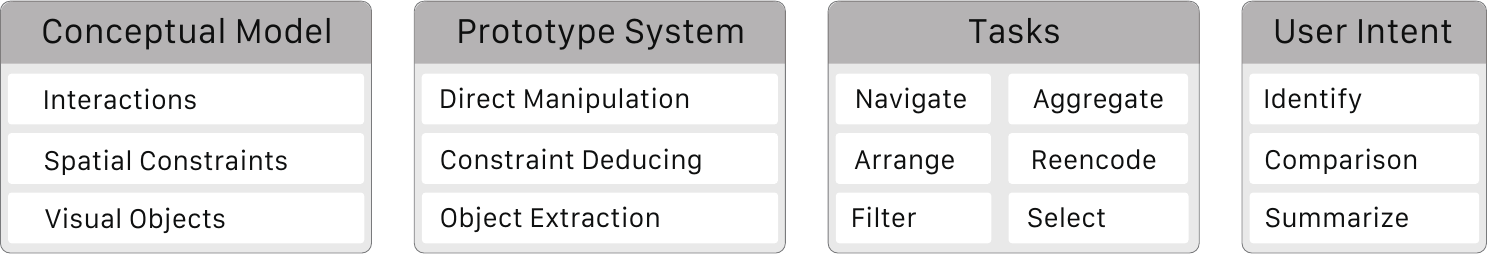}
    \caption{We present a spatially-constrained conceptual model. Building upon this conceptual model, we have implemented a prototype system. These interactions can facilitate various user interaction tasks and accommodate diverse user intentions.
   }
    \label{fig:whole_overview}
\end{figure*}

From this classification, we can observe that changes triggered by many types of interactions can be described as spatial changes.
Therefore, the spatial constraint method can model numerous interactions and the resulting spatial changes.
To support these spatial-related interaction tasks, we propose a spatial constraint method that models the positions of visual objects.
Direct manipulation of visual objects allows the constraints to guide them into a new stable state, resulting in smooth transitions during interactions.
The spatial changes support users' tasks, such as comparison and identification.

As shown in \autoref{fig:whole_overview}, we present a conceptual framework in the form of a spatial constraint model that elucidates spatial relationships among visual objects through spatial constraints. Specifically, it clarifies the rationale behind the existing positions of visual objects and anticipates their transformations when subjected to changes. Visual objects characterized by spatial constraints are suitable for accommodating spatial interactions. Building upon this conceptual foundation, we have developed a prototype system. This system extracts existing visualizations, including the visual objects within them, and deduces latent spatial constraints among these entities. Furthermore, it facilitates the direct manipulation of visual objects, coordinate axes, and spatial constraints based on these inferred constraints.
These direct manipulations empower users with data analysis tasks to execute a diverse array of actions aimed at addressing a spectrum of user intents.

\begin{figure*}[!ht]
    \centering
    \includegraphics[width=\textwidth]{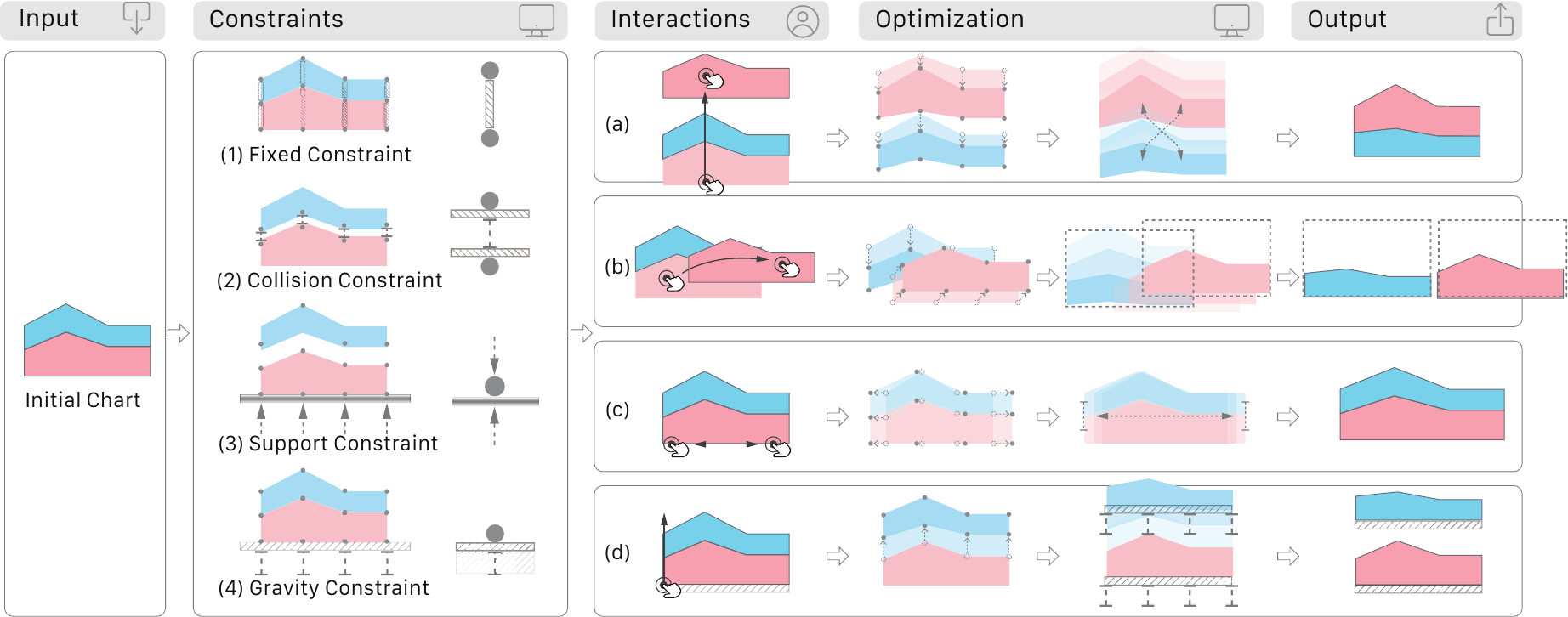}
    \caption{A chart can be constructed using four fundamental constraints: fixed, collision, support, and gravity constraints. These constraints facilitate the direct manipulation of the visualization, encompassing (a) and (b) the manipulation of visual objects, (c) manipulation of axes, and (d) manipulation of constraints.
    Spatial constraints are translated into forces that prompt positional changes in visual objects, analogous to how forces impact the positions of physical objects. The resultant converged state enables a variety of interactive tasks, including rearrangement, deletion, and navigation.
   }
    \label{fig:pipeline}
\end{figure*}

As shown in \autoref{fig:pipeline}, spatial constraints are established for the visualization, allowing users to manipulate visual objects, axes, and constraints through intuitive actions.
The manipulated control points and positions of constraints serve as inputs for the optimization process, which utilizes encoded forces to alter the positions of visual objects and attain a new stable layout.
The concept of spatial constraints and the four types of atomic constraints are explored in \autoref{section:constraints}, while \autoref{section:manipulation} delves into direct interactions with the visualization supported by spatial constraints.

 \section{Spatial Constraints for Visualization}
\label{section:constraints}

We introduce a spatial constraint model to depict the positioning of visual objects within visualizations. We provide an overview of spatial constraints frequently employed in visualizations and establish definitions for \textbf{atomic constraints} to model spatial layouts. We illustrate the applicability of atomic constraints in representing typical visualizations through examples.

\subsection{Modeling Visualization Layouts with Forces}
\label{sec:vis_force_case}

\begin{figure}[!htb]
    \centering
    \setlength{\belowcaptionskip}{10px}
    \includegraphics[width=.9\columnwidth]{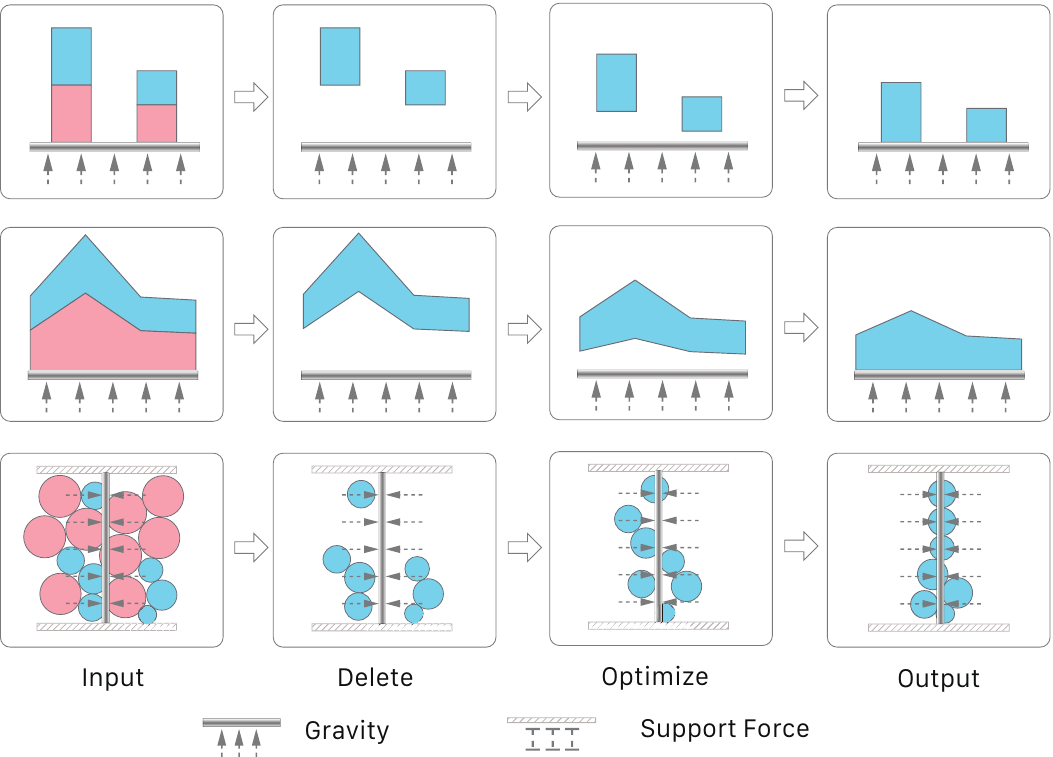}
    \caption{The position of the existing visual objects will be updated after some visual objects are removed.}
    \label{fig:filter_circle}
\end{figure}

This section introduces a conceptual framework that models visual objects with constraints, using several illustrative scenarios. As depicted in Figure \ref{fig:filter_circle} (a), each bar within the chart can be treated as an area-type object with four control points. When implementing the bar chart programmatically, the position and shape of each bar are determined using predefined algorithms. For instance, if the chart is constructed using a tool like D3~\cite{bostock2011d3}, the programmer assigns attributes (such as \texttt{x}, \texttt{y}, \texttt{width}, and \texttt{height}) to each \texttt{<rect>} element.
However, in situations where certain bars, such as the pink bars, are removed, the programmer must implement a function to readjust the positions of the remaining bars, aligning them with the $x$-axis. Achieving fluid transitions in such cases requires additional significant effort.

In our approach, bars can be conceptualized as physical objects stacked on a supporting surface (ground). All objects are influenced by gravity, naturally descending towards the ground. In this context, the $x$-axis serves as the supporting ground, facilitating intuitive and automated movement of visual objects. As depicted in Figure \ref{fig:filter_circle}, when the pink bars are removed, the blue bars naturally descend to rest upon the $x$-axis, aided by the ``hold-up'' force exerted by the $x$-axis. Consequently, the final spatial arrangement of visual objects results in a simplified bar chart.
Furthermore, beyond external forces impacting visual objects, there are internal forces preserving the shapes of these objects. For instance, the fixed distance between the bottom-left and top-left corner points is maintained both vertically and horizontally.

\autoref{fig:filter_circle} depicts a stacked area chart and a bubble chart. After the removal of the pink area in the stacked area chart, the blue area descends to the "ground". Within the stacked area chart, each point is influenced by multiple horizontal positions (i.e., ticks). Additionally, there are collision forces among the visual objects in the vertical direction to prevent the blue area from overlapping with the pink area. In the bubble chart, collision forces also exist among the bubbles. Simultaneously, all the points are drawn towards the center in the horizontal direction. In these examples, the control points are subject to three types of constraints:
\begin{itemize}
  \item \textbf{Environmental Constraints}: In the bar chart and area chart depicted in \autoref{fig:filter_circle}, each visual object is influenced by both gravity and "hold-up" forces in the vertical direction. These forces are referred to as the \textbf{gravity} and \textbf{support} constraints, respectively.
  
  \item \textbf{Inter-object Constraints}: Vertical collision forces come into play among visual objects, such as the stacked blue and pink bars in \autoref{fig:filter_circle}. These \textbf{collision} constraints ensure that the blue bars are stacked atop the pink bars, preventing overlap. Additionally, for point-like visual objects, collision forces maintain separation between circles.
  
  \item \textbf{Intra-object Constraints}: Visual objects also have \textbf{fixed} relationships among their internal points to preserve their shapes. For instance, fixed vertical distances between control points (e.g., between the top-left and bottom-left corners) maintain a constant bar height.
\end{itemize}

\subsection{Control Points}

Three types of visual objects exist: points (\inlinegraphics{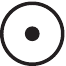}), lines (\includegraphics{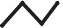}), and areas (\includegraphics{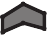}). Control points serve as the fundamental units for these visual objects. These objects manifest as individual control points, connected sequences of control points forming lines, and enclosed regions defined by control points.
In the context of SVG (Scalable Vector Graphics), control points for lines and areas can be extracted by parsing the endpoints of each segment from an SVG path. For instance, in a bar chart, each bar is defined by its four corner points, which function as control points. Similarly, in a circular chart, each circle is represented as a point with an associated radius.
In a stacked area chart, points encompass the areas of each visual object. Presently, our focus is on 2-D visualization, where a control point occupies a position defined by two dimensions. Formally, a control point is represented as $$P_i = \{(x, y), r\},$$ where $x$ and $y$ denote coordinates along the horizontal and vertical axes in a Cartesian coordinate system. The value of $r$ represents the point's radius.
For line-type and area-type visual objects, the radii of their control points are set to $0$, while a point-type object may possess a non-zero radius. This distinction in radii contributes to the nuanced representation of these various visual elements.

\subsection{Atomic Constraints}

As summarized in subsection~\ref{sec:vis_force_case}, our analysis reveals four fundamental atomic constraints: gravity, support, collision, and fixed constraints. The subsequent paragraphs provide the formal definitions for each of these constraints.

\begin{wrapfigure}{r}{0.16\columnwidth}
    \begin{center}
      \vspace{-0.09\columnwidth}
      \includegraphics[width=0.16\columnwidth]{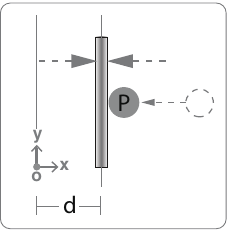}
      \vspace{-0.03\columnwidth}
    \end{center}
\end{wrapfigure}
\textbf{Gravity constraints} induce an attractive force towards a control point $P$ from a specific position, either horizontally or vertically. Formally, a gravity constraint for a control point $P$ in the $x$ direction is established by minimizing $$f_g\left(x - d\right),$$ where $f_g$ increases as the magnitude of $|x - d|$ increases.
To ensure the continuity and differentiability of gravity constraints for optimization purposes, we define $$f_g \left( t\right) = (t)^2,$$ which represents the simplest function that maintains continuity and differentiability. The accompanying diagram on the right provides an illustrative instance of gravitational attraction towards the control point in the $x$-direction.

\begin{wrapfigure}{r}{0.16\columnwidth}
    \begin{center}
      \vspace{-0.03\columnwidth}
      \includegraphics[width=0.16\columnwidth]{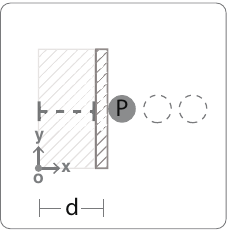}
      \vspace{-0.03\columnwidth}
    \end{center}
\end{wrapfigure}
\textbf{Support constraints} exert controlled influence on the coordinates of a given control point $P$, either compelling them to exceed or remain below specific thresholds along the horizontal or vertical axes. 
An instance is enforcing a point's positioning above a designated threshold. 
This can be observed in \autoref{fig:filter_circle}, where support constraints are enforced on the bars from $x$-axis.
Mathematically, the support constraint for a control point $P$ is defined as
$$x \pm \xi = d,$$
where $\xi$ represents a non-negative slack variable. This constraint ensures that $\xi \geq 0$ in all cases. When the positive sign is selected for $\pm$, as in $$x + \xi = d,$$
control point $P$ is positioned to the left of the distance $d$.

\begin{wrapfigure}{r}{0.16\columnwidth}
  \begin{center}
    \vspace{-0.03\columnwidth}
    \includegraphics[width=0.16\columnwidth]{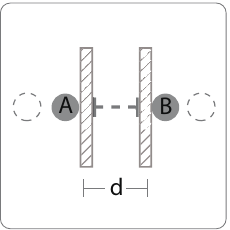}
    \vspace{-0.04\columnwidth}
  \end{center}
\end{wrapfigure}
\textbf{Collision constraints} pertain to the spatial relationship between control points. Two types of collision constraints exist: the collision of control points along the vertical or horizontal axes and the collision relationship among point-type visual objects. The former guarantees specific positions relative to each other, positioning control point $A$ to the left, right, above, or below control point $B$.
Mathematically, the constraint in the x-direction is expressed as
$$x_{b} - x_a - d \pm \xi = 0,$$
When the $\pm$ sign is negative, thus yielding
$$x_{b} - x_a - d - \xi = 0,$$
as illustrated in the right figure, Point $B$ is situated to the right of Point $A$, with a minimum distance of $d$ between them.

\begin{wrapfigure}{r}{0.16\columnwidth}
    \begin{center}
      \vspace{-0.06\columnwidth}
      \includegraphics[width=0.16\columnwidth]{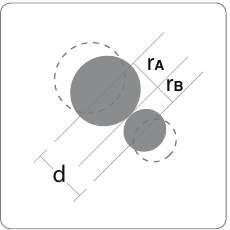}
      \vspace{-0.08\columnwidth}
    \end{center}
\end{wrapfigure}
The collision constraints among point-type visual objects prevent overlapping of points within a 2D space.
This constraint guarantees that the distance between two points exceeds the sum of their radii. The collision constraint between two points, $P_a$ and $P_b$, can be expressed as
$$\ d\left(P_a, P_b\right) = \sqrt{(x_a - x_b)^2 + (y_a - y_b)^2} \geq r_a + r_b.$$

\begin{wrapfigure}{r}{0.16\columnwidth}
    \begin{center}
      \vspace{-0.04\columnwidth}
      \includegraphics[width=0.16\columnwidth]{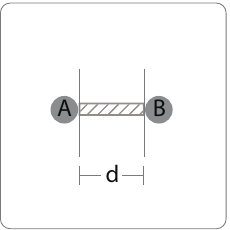}
      \vspace{-0.12\columnwidth}
    \end{center}
\end{wrapfigure}
\textbf{Fixed constraints} are related to the fixed distance of control points inside a visual object.
For example, the fixed-distance between corner points of a bar.
Formally, a fixed constraint for points $A$ and $B$ in the x-direction is presented as $$(x_a - x_b - d) = 0.$$

\subsection{Modeling Visualization Layouts with Atomic Constraints}

\begin{figure}[!ht]
  \centering
  \includegraphics[width=.8\columnwidth]{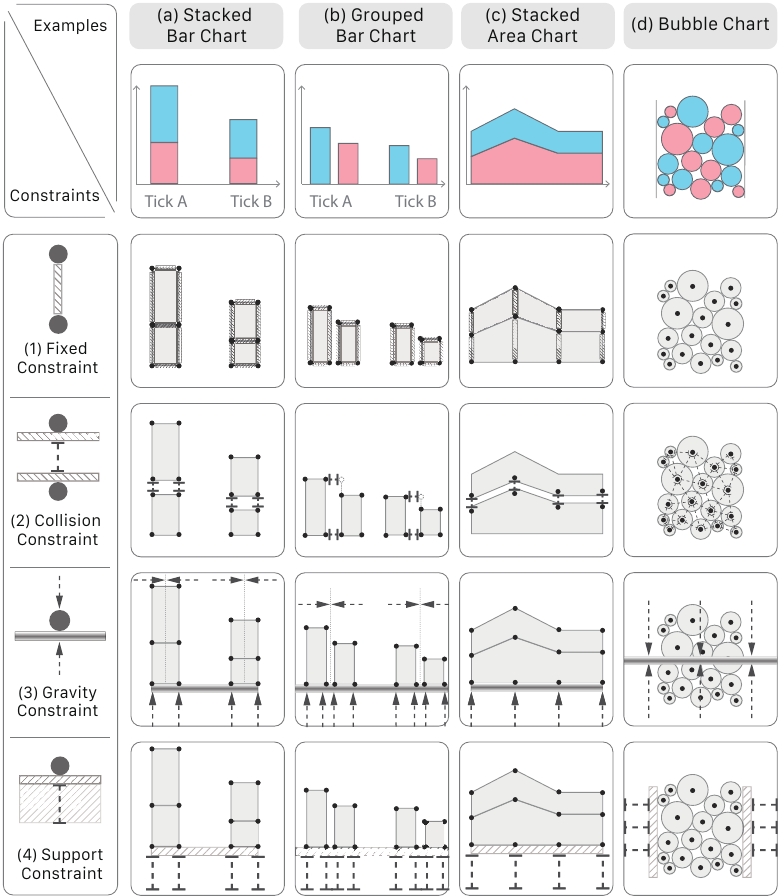}
  \caption{Constraints for exemplary visualizations. Each column corresponds to a visualization type, and each row represents a specific category of atomic constraints.}
  \label{fig:chart_constraint}
\end{figure}

This subsection offers an introduction to the methodology of representing common visualizations using atomic constraints. \autoref{fig:chart_constraint} showcases a matrix that includes four common visualizations, each accompanied by its corresponding atomic constraints.

Within the conceptual framework, our spatial constraint model can be built upon control points, integrating a range of constraint types, thus facilitating versatile representations. Commencing with visualizations articulated in control point format within a two-dimensional Cartesian coordinate system, we employ a set of visualizations as case studies. These encompass both categorical and quantitative axes.

\begin{itemize}
\item \textbf{Gravity constraints:}
Coordinate axes can apply gravity constraints to their associated positions. For example, in a line chart or scatter plot, the positions of control points are defined by the data they represent along their respective coordinates. Gravity constraints are solely relevant to line charts and scatter plots, with no involvement of collision or fixed constraints. As the coordinate axes are rescaled, the visual objects within these charts experience scale adjustments accordingly.

\item \textbf{Fixed constraints:} Slightly more complex visualizations incorporate fixed constraints, as seen in bar charts and area charts.
Fixed constraints can align with data mappings or pre-established configurations. For instance, in the context of a bar chart, the height corresponds to data values, while the width represents pre-defined settings.

\item \textbf{Support constraints:}
Moreover, in the case of stacked visualizations along an axis, gravity and support forces can be utilized to guarantee the vertical alignment of content along the x-axis (assuming we are discussing visualizations stacked along the x-direction).

\item \textbf{Collision constraints:}
Interactions among visual objects lead to compression and collision effects. For instance, take a stacked area chart, where a collision relationship exists between the visual objects both above and below, requiring collision constraints.
In a grouped bar chart, collision constraints are relevant to the left and right visual objects within a group.

\end{itemize}

Expanding on these insights, we illustrate the process of constraint inference through various representative examples. Specifically, for the purpose of demonstration, we offer instances involving stacked bar charts, grouped bar charts, stacked area charts, and bubble charts. Following this, we broaden our analysis to encompass basic bar charts, area charts, line charts, and scatter plots.
All these instances are rooted in visualizations within a two-dimensional Cartesian coordinate system.

\begin{itemize}
\item \textbf{Stacked bar chart}.
\autoref{fig:chart_constraint} (a) displays a stacked bar chart alongside its four types of atomic constraints.
Each bar adheres to fixed constraints along all four edges.
Vertical collision constraints exist among the stacked bars within a group, ensuring their non-overlapping alignment.
Simultaneously, all bars experience a horizontal gravitational force that pulls them towards the center position of the tick marks (such as ``Tick A'' for the left bars).
Moreover, all control points are subjected to gravitational and supportive constraints originating from the $x$-axis, designated as the \textbf{baseline axis}.
Through the collective interplay of these four constraint types, the visual objects assume their respective positions.

\item \textbf{Grouped bar chart}.
As depicted in \autoref{fig:chart_constraint} (b), the fixed constraints, gravity constraints, and support constraints applicable to a grouped bar chart closely mirror those of a stacked bar chart.
Within a single group, horizontal collision constraints arise among the bars.
The fundamental divergence between a grouped bar chart and a stacked bar chart pertains solely to the orientation of the collision constraints.

\item \textbf{Stacked area chart}.
A stacked area chart typically portrays the temporal evolution of multiple data series.
As depicted in \autoref{fig:chart_constraint} (c), the control points forming a vertical line in a visual object are governed by fixed constraints, ensuring a constant height at a specific x-position.
Each point's x-position is subject to horizontal gravity constraints.
Analogous to a stacked bar chart, gravity constraints and support constraints originating from the baseline axis are also applicable.
Furthermore, collision constraints in the vertical direction prevent visual objects from overlapping.

\item \textbf{Bubble chart}.
\autoref{fig:chart_constraint} (d) illustrates a bubble chart, where bubbles experience compression from the left and right sides while being drawn towards the horizontal line.
The absence of overlap among bubbles is due to collision constraints.
A vertical gravity constraint originates from the center, while support constraints are present on the left and right sides.

\item \textbf{Other visualizations.}
The constraints can also be applied to model other prevalent visualizations. For instance, a simple bar chart exhibits a subset of the constraints found in a stacked bar chart, with the exception of collision forces.
Regarding line charts and scatter plots, collision constraints are not applicable to visual objects within them.
Gravity constraints in both horizontal and vertical directions can effectively underpin the modeling process.

\end{itemize}

In stacked/grouped bar charts and stacked area charts, collision constraints exclusively pertain to the control points of distinct visual objects sharing the same tick value on the $x$-axis; this subset of objects is termed the collision group.
To illustrate, in \autoref{fig:chart_constraint}, stacked or grouped bars occupying identical tick positions on the $x$-axis, or points aligned along the same vertical line within the stacked area chart, form part of a collision group.
The control points within a collision group maintain an order and direction, denoted as the collision order and collision direction, respectively.
For instance, the stacked area chart presented in \autoref{fig:chart_constraint} (c) exhibits a collision order: the blue visual element is positioned above the red one, with a vertical collision direction.

We utilize these examples to demonstrate the efficacy of atomic constraints.
Such constraints can stabilize control points in their present locations and facilitate enhanced interactivity, a topic elaborated upon in ~\autoref{section:manipulation}.

\subsection{Construction of Constraints for Existing Charts}

We present a heuristic algorithm for constructing constraints in an existing visualization based on the analysis of prevalent design patterns.
Initially, we extract the control points from the given chart.
Subsequently, we establish the \textbf{visual object set}, which constitutes a collection of visual objects of the same category.
In the case of area objects (such as bars and areas), the constraint construction involves detecting the baseline axis, calculating collision constraints, and determining fixed constraints.
Regarding point objects (like points and bubbles), we incorporate a point collision detection mechanism.
For line objects (including line charts), since there typically exist no collision constraints among visual elements (as lines are usually not stacked together), we can assign gravity constraints to maintain their current positions.

\textbf{Control point extraction.}
We initiate the process by utilizing a visualization in SVG format to extract the control point positions.
In our methodology, a \texttt{<circle>} element is interpreted as a point visual object.
A line or an open path in the SVG corresponds to a line visual object.
Closed paths and rectangles are parsed as area visual objects.
The absolute coordinates of each control point are calculated.
Subsequently, we compile a roster of visual objects along with their corresponding control points.
A bar encompasses four control points located at its corners.
In contrast, each bubble constitutes an object with a solitary control point positioned at its center.
For area objects, control points are positioned at the termini of each segment within the \texttt{<path>} element of the SVG.

\textbf{Visual object set extraction.}
A visual object set comprises a group of visual objects employed in establishing constraints.
A visualization encompasses a series of data items depicted by a set of visual objects.
An area object set facilitates the creation of diverse chart types, such as stacked and basic area charts, as well as stacked, grouped, and basic bar charts.
For instance, a set of uniformly spaced bars forms a bar chart.
Point sets are employed in crafting scatter plots and bubble charts.
A singular line or an assemblage of lines can be harnessed to craft a line chart.

\begin{figure}[htbp]
    \centering
    \setlength{\belowcaptionskip}{-10px}
    \includegraphics[width=\columnwidth]{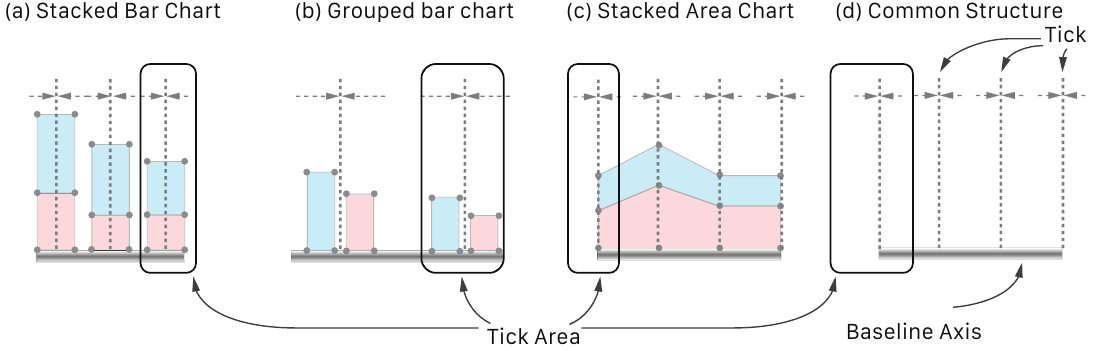}
    \caption{Baseline axes and ticks of bar charts and area charts.}
    \label{fig:bar_area_deducing}
\end{figure}

\textbf{Baseline axis detector.}
For bar charts and area charts, a baseline axis is established, which enforces gravity and support constraints on the visual objects.
As demonstrated in \autoref{fig:bar_area_deducing}, a shared configuration (d) featuring a baseline axis showcases multiple parallel tick lines perpendicular to the baseline.
By analyzing the positions of control points and visual objects, we identify the baseline.
In both stacked and simple bar charts, the centers of bars are uniformly aligned along the baseline axis, simplifying the computation of tick positions via these visual object centers.
Similarly, in stacked area charts, control points oriented along the baseline axis are uniformly distributed, enabling us to designate these equidistant positions as ticks.
In the scenario of grouped bar charts, the intermediate positions on the scale do not exhibit perfect uniformity. Consequently, we ascertain these positions in conjunction with the arrangement of text (which is uniformly distributed) along the axes.

\begin{figure}[htbp]
  \centering
  \includegraphics[width=\columnwidth]{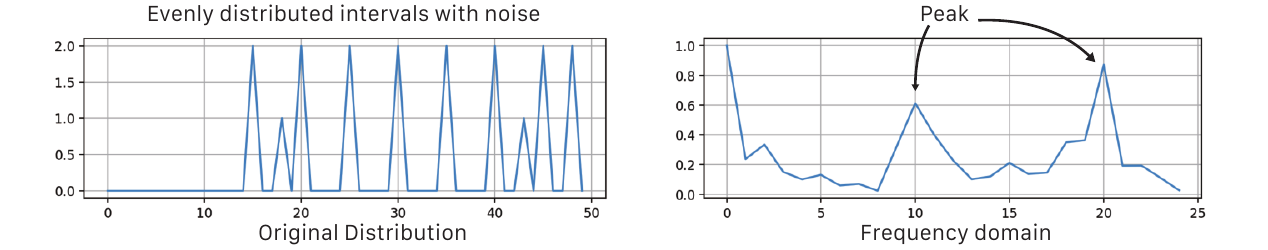}
  \caption{Evenly distributed intervals can be extracted using the peaks in the frequency domain even with some noise.}
  \label{fig:evenly_distributed}
\end{figure}

The central step in establishing the baseline axis involves assessing the uniform distribution of visual objects, control points, and text along the vertical or horizontal dimension.
We derive their positions in the x- and y-directions, followed by a transformation into the frequency domain using Fourier analysis.
Illustrated in \autoref{fig:evenly_distributed}, the identification of evenly spaced intervals is facilitated by identifying peaks in the frequency domain.
As shown in \autoref{fig:evenly_distributed}, positions are projected with a 5-pixel interval, accompanied by some degree of noise.
By detecting peaks within the frequency domain, we can extract distribution intervals.
This interval corresponds to a wavelength within the frequency domain, and its calculation is as follows:
$interval = \frac{50}{10} = 5.$
By employing the computed interval to filter out noise, we deduce the positions of ticks.
Upon tick detection, the baseline direction is established.
With the identification of the baseline axis direction, the baseline position can be readily determined using the positions of visual objects along the perpendicular direction.
Subsequently, the support constraints and gravity constraints are ascertained.

\textbf{Calculation of collision constraints.}
Collision constraints solely come into effect for control points of distinct visual objects situated within the same tick area, as depicted in \autoref{fig:bar_area_deducing}.
We ascertain the absence of overlapping by evaluating both the x and y directions of the control points.
In instances where a group of visual objects within a tick area exhibit non-overlapping alignment in a specific direction, we conclude that these elements possess collision constraints with one another.
For instance, in a grouped bar chart, collision constraints are oriented parallel to the baseline axis to ensure bars align in parallel along it.
In contrast, stacked area and stacked bar charts impose collision constraints perpendicular to the baseline axis to ensure elements are orthogonally stacked.

\textbf{Calculation of fixed constraints.}
Fixed constraints are confined to control points situated within the boundaries of the same visual objects and tick intervals. As depicted in \autoref{fig:bar_area_deducing}, we can enforce fixed constraints on visual objects that fall within a shared tick interval. In cases where a visual object is entirely contained within a tick interval, such as a bar in a bar chart, we establish fixed constraints on the two vertices of an edge along each of the four sides, thereby preserving the edge's positioning. In situations where elements extend across multiple tick intervals, fixed constraints are applied to the same element within the corresponding tick interval. For instance, in the context of an area chart, fixed constraints are set on two points at the same x-position perpendicular to the axis.

\textbf{Point collision detector.}
We propose a point collision detection algorithm designed to ascertain whether a collection of point objects is subject to collision constraints. When a series of circles are closely situated, the separation between pairs of circles approaches the summation of their radii, resulting in $Distance(A, B) - r_a - r_b \sim 0$.
Our collision detection algorithm computes the distribution of $Distance(A, B) - r_a - r_b$, with the expected minimum value being approximately 0 if collision constraints exist. In scenarios involving numerous interconnected pairs of circles, the FloodFill algorithm is employed to identify a continuous area of interconnected circles, originating from a specified circle.

For visual object groupings devoid of collision constraints, such as line charts and scatter plots, the gravity constraints can be aligned with the present positions of control points along both the x- and y-axes. In such visualizations, axis scaling can be adjusted to modify the gravity constraints on the control points.

It is noteworthy that the algorithms presented in this section serve as inference algorithms, grounded in prevalent implementation approaches for visualizations. The algorithm is founded on vector visualization and encompasses several established principles, such as aligning multiple labels horizontally or vertically along the axes. It encompasses a range of common visualizations, including bar charts (stacked, grouped, simple), area charts (stacked, overlapped, simple), bubble charts (including collision detection), scatter plots, and line charts.
Nevertheless, this inference method does not represent the sole solution, as alternative reverse engineering approaches~\cite{savva2011revision, poco2017reverse} for extracting visualization results can also be equivalently expressed as outcomes derived from the inference process outlined in this method.

\begin{algorithm}[!htb]
    \caption{Optimization Process: the calculation of the position of control points.}
    \label{alg:optimization}
    \begin{algorithmic}[1]
    \Require 
    \Statex Control points list $P_i$, with position $(P_i.x, P_i.y)$, velocity $(P_i.v_x, P_i.v_y)$, and radius $(P_i.r)$, $P_i.v_x = 0$, $P_i.v_y = 0$;
    \Statex Gravity constraints list $G_i$, $i = 1, 2 \ldots N_g$, $G_i = \{P, d, dim\}$; \Comment{$P$ is the position; $dim$ is the coordinate direction of $G_i$, i.e., $dim \in \{x, y\}$}
    \Statex Support constraints list $S_i$, $i = 1, 2 \ldots N_s$, $S_i = \{P, d, dim, op\}$; \Comment{$op \in \{\leq, \geq\}$.}
    \Statex Fixed constraints list $F_i$, $i = 1, 2 \ldots N_f$, $F_i = \{P_{1}, P_{2}, d, dim\}$; \Comment{$d$ is the distance.}
    \Statex Collision I constraints list $L_{i}$, $i = 1, 2 \ldots N_l$, $L_i = \{P_{1}, P_{2}, d, dim\}$; %
    \Statex Collision II constraints list $C_i$, $i = 1, 2 \ldots N_c$, $C_i = \{P_{1}, P_{2}, d\}$; %
    \Statex alpha; \Comment{The strength of current iteration.}
    
    \For{ $i = 1;\ i \leq N_g;\ i \gets i + 1$}\Comment{Handle gravity constraints.}
    \State $dim = G_i.dim$\Comment{$dim \in \{x, y\}$}
    \State $\epsilon \gets G_i.P[dim] + G_i.P.v[dim] - G_i.distance$
    \State $G_i.P.v[dim] \gets G_i.P.v[dim]  -\epsilon * alpha$; 
    \EndFor

    \For{ $i = 1;\ i \leq N_s;\ i\gets i + 1$}\Comment{Handle support constraints.}
    \State $dim = S_i.dim$
    \If{not $S_i.P[dim] + S_i.P.v[dim]\ S_i.op\ S_i.d$}\Comment{$S_i.op \in \{\leq,\geq\}$}
    \State $S_i.P[dim] \gets S_i.d$
    \State $S_i.P.v[dim] \gets 0$; 
    \EndIf
    \EndFor

    \For{ $i = 1;\ i \leq N_f;\ i\gets i + 1$}\Comment{Handle fixed constraints.}
    \State $dim = F_i.dim$; $d \gets F_i.d$;  $P_1 \gets F_i.P_1$; $P_2 \gets F_i.P_2$
    \State $\epsilon \gets P_{1}[dim] + P_{1}.v[dim] - P_{2}[dim] - P_{2}.v[dim] - d$
    \State $P_{1}[dim] \gets P_{1}[dim] - \epsilon * 0.5$;  $P_{2}[dim] \gets P_{2}[dim]  + \epsilon * 0.5$
    \State $P_{1}.v[dim], P_{2}.v[dim] \gets (P_{1}.v[dim] + P_{2}.v[dim])/2$
    \EndFor
    \For{ $i = 1;\ i \leq N_l;\ i\gets i + 1$}\Comment{Handle collision I constraints.}
    \State $dim \gets L_i[dim]$; $P_1 \gets L_i.P_1$; $P_2 \gets L_i.P_2$
    \State $\epsilon \gets P_{1}[dim] + P_{1}.v[dim] - P_{2}[dim] - P_{2}.v[dim] - L_i.d$
    \If{$\epsilon < 0$}
    \State $P_{1}[dim] \gets P_{1}[dim]  -\epsilon * 0.5$; $P_{2}[dim] \gets P_{2}[dim]  +\epsilon * 0.5$
    \State $v_{avg} \gets P_{1}.v[dim] + P_{2}.v[dim]$
    \State $P_{1}.v[dim] \gets v_{avg}$; $P_{2}.v[dim] \gets v_{avg}$
    \EndIf
    \EndFor
    
    \For{ $i = 1;\ i \leq N_c;\ i\gets i + 1$}\Comment{Handle collision II constraints.}
    \State $P_1 \gets C_i.P_1$; $P_2 \gets C_i.P_2$
    \State $d_x \gets (P_{1}.x + P_{1}.v_x - P_{2}.x - P_{2}.v_x)$
    \State $d_y \gets (P_{1}.y + P_{1}.v_y - P_{2}.y - P_{2}.v_y)$
    \State $d \gets \sqrt{(d_x)^2 + (d_y)^2}$
    \State $\epsilon \gets d - N_i.d$
    \If{$\epsilon < 0$}
    \State $bias \gets (P_{1}.radius) ^ 2 / ((P_{1}.radius) ^ 2 + (P_{2}.radius) ^ 2)$
    \State $P_{1}.v_x \gets P_{1}.v_x - \epsilon * d_x / d * bias$
    \State $P_{1}.v_y \gets P_{1}.v_y - \epsilon * d_y / d * bias$
    \State $P_{2}.v_x \gets P_{2}.v_x - \epsilon * d_x / d * (1- bias)$
    \State $P_{2}.v_y \gets P_{2}.v_y - \epsilon * d_y / d * (1 - bias)$
    \EndIf
    \EndFor
    \For{ $i = 0; i < N_p; i\gets i + 1$}\Comment{Update control points' position.}
    \State $P_i.x \gets P_i.x + P_i.v_x$; $P_i.y \gets P_i.y + P_i.v_y$
    \EndFor
    \end{algorithmic}
\end{algorithm}

\subsection{Optimization Process}

The interaction process involves three key stages: dragging, dropping, and optimization, as depicted in \autoref{fig:pipeline}.
Spatial constraints guide visual objects towards a new equilibrium.
To make the optimization process interactive and continuous, we employ physical forces to encode these constraints.
Once visual objects, control points, and constraints are extracted from the visualization, we convert the constraint conditions into forces that facilitate spatial transformations of the control points.
Each visual object is linked to an ordered sequence of control points, collectively defining point, line, or area-type visual entities based on their arrangement.
Every control point possesses attributes of position and velocity. During each iteration, these control points update their positions and velocities, resulting in the rendering of a new visual object based on the revised control point positions.
We leverage D3's force-directed simulation~\cite{bostock2011d3}, tailoring the forces to accommodate spatial constraint integration.
Further details are provided in Algorithm~\ref{alg:optimization}.
The computational complexity of a force-directed layout is $O(k * (n^2))$, where $k$ denotes the number of iterations, and $n$ signifies the count of control points.
Except for circle collisions, constraints operate solely along horizontal or vertical axes.
Consequently, the influence of each control point on others is limited, curbing the magnitude of $n$ in the complexity equation.
For circle collisions, a quad-tree division is employed before computation, reducing the collision complexity to $O(kn\ log(n))$.
In a visualization featuring 1,000 control points (e.g., a bar chart with 250 bars), the optimization process can converge within a matter of seconds.

\section{Manipulation Supported by Constraints}
\label{section:manipulation}

\begin{figure*}[!ht]
  \centering
  \includegraphics[width=\textwidth]{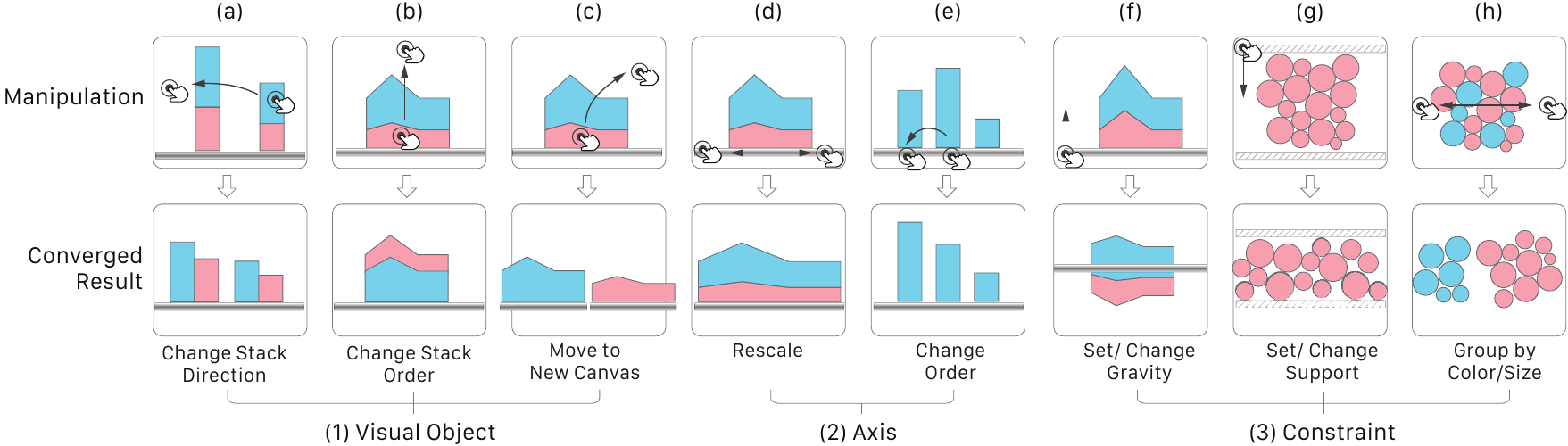}
  \caption{
    Our approach empowers users to directly manipulate visual objects, axes, and constraints. It accommodates a wide range of interaction tasks within visualizations.
  }
  \label{fig:manipulate_classification}
\end{figure*}

As depicted in \autoref{fig:manipulate_classification}, direct manipulations can be classified into three distinct types based on their targets: visual objects, axes, and constraints. These manipulations are respectively referred to as object-level, axis-level, and constraint-level manipulations.

\subsection{Manipulating Visual Objects}
\label{sec:manipulate_object}

Object-level manipulations directly alter the positions of visual objects.
User intent might be implicitly embedded within object manipulations.
Manipulations involving distinct directions, distances, and velocities convey diverse user intentions.
For instance, dragging visual objects along the collision direction implies adjustments to stacking or grouping order, while moving a visual object beyond the current canvas suggests transferring it to a new canvas.

\textbf{Changing the stacking direction.}
A set of visual objects can be stacked either along the $x$-direction (e.g., in a grouped bar chart) or the $y$-direction (e.g., in a stacked bar chart).
Examples of changing the stacking direction include transforming a grouped bar chart into a stacked bar chart, converting a stacked bar chart into a grouped bar chart, or altering a stacked area chart into an overlapped area chart or vice versa.
The alteration of stacking direction involves reconfiguring collision constraints to align with the new direction.
For example, in \autoref{fig:manipulate_classification} (a), when collision constraints along the $y$-direction shift to the $x$-direction, the newly introduced collision constraints cause the blue bars to spread horizontally from the red bars.
At the same time, the downward gravitational force affects the descent of the blue bars.
Transforming a stacked bar chart into a grouped bar chart facilitates intra-group bar comparisons, while converting a grouped bar chart into a stacked bar chart aids in depicting totals.

\textbf{Changing the stacking order.}
As demonstrated in \autoref{fig:manipulate_classification} (b), the user performs a drag action, moving the pink bar above the blue bar.
This action implies an intention to alter the stacking order of these two visual objects.
We subsequently recompute the collision order based on the updated positions of these visual objects.
Specifically, within each tick area, the horizontal position of the pink bar is adjusted to be higher than that of the blue bar.
Consequently, the gravitational force pushes the blue bar downward towards the $x$-axis, resulting in the pink bar descending and being stacked atop the blue bar.

\textbf{Moving to a new canvas.}
When a visual object (or a group of visual objects) is dragged beyond the canvas boundaries (\autoref{fig:manipulate_classification} (c)), a new canvas is generated to accommodate the relocated visual object(s).
The spatial constraints from the original canvas are inherited by the new canvas.
Manipulation and optimization are carried out independently for the visual objects within each canvas.
Visual objects can be moved between existing canvases by dragging.
Rapidly dragging and dropping visual objects outside the canvas will result in the deletion of these objects.

\subsection{Manipulating Axes}
\label{sec:manipulate_axes}

Axis-level manipulation is founded on the alteration of axis-related constraints driven by user interactions.
These constraints encompass gravity, support, and fixed constraints, all aligned with the axis direction.
Manipulations on the axis lead to rescaling and reordering visual objects.
Based on the encoded data attributes, axes are categorized into two types: those with continuous attributes (quantitative and temporal) and those with discrete attributes (categorical).
The determination of axis type is contingent upon the labels present on the axis ticks. An axis is classified as continuous if its text labels can be converted to numeric values or time stamps;
otherwise, it is interpreted as a discrete axis.
Manipulations executed on the axis engender changes in its scale or order.
For continuous axes, users can effect rescaling or reordering by magnifying (via scrolling or pinching) the axis or dragging its ticks.
On discrete axes, tick rearrangements can be accomplished through tick dragging, and rescaling can be achieved by zooming (e.g., altering the widths of bars).
Correspondingly, as the scale or order undergoes modification, the associated constraints evolve in tandem.
These constraints also act as drivers for the convergence of control points to new positions.

\textbf{Rescaling axes.}
As shown in \autoref{fig:manipulate_classification} (d), users can zoom in on a continuous axis to change the scale of the axis.
In our method, users can zoom in on a continuous axis in three ways, namely, by pinching, scrolling, and dragging a tick label.
The scale of the axis changes according to the zooming rate.
According to the changed scale, our model updates the gravity, support, and fixed constraints of the control points in the axis direction.

\textbf{Reordering axes.}
If the axis is discrete (e.g., categorical), users can reorder the axis directly by dragging the ticks, as shown in \autoref{fig:manipulate_classification} (e).
After dragging the ticks, the order of the ticks is recalculated according to the ticks' new positions.
The gravity constraints of a specific tick on the axis are changed accordingly.
Moreover, as dragging each tick to sort the ticks is time-consuming, we also implement a sort function; by right-clicking the axis, the ticks can be sorted according to selected visual objects' attributes (e.g., width, height, left, right, or color).
For example, we can sort the bars of a bar chart according to the height of the bars.

\begin{figure}[htb]
    \centering
    \setlength{\belowcaptionskip}{-10px}
    \includegraphics[width=\columnwidth]{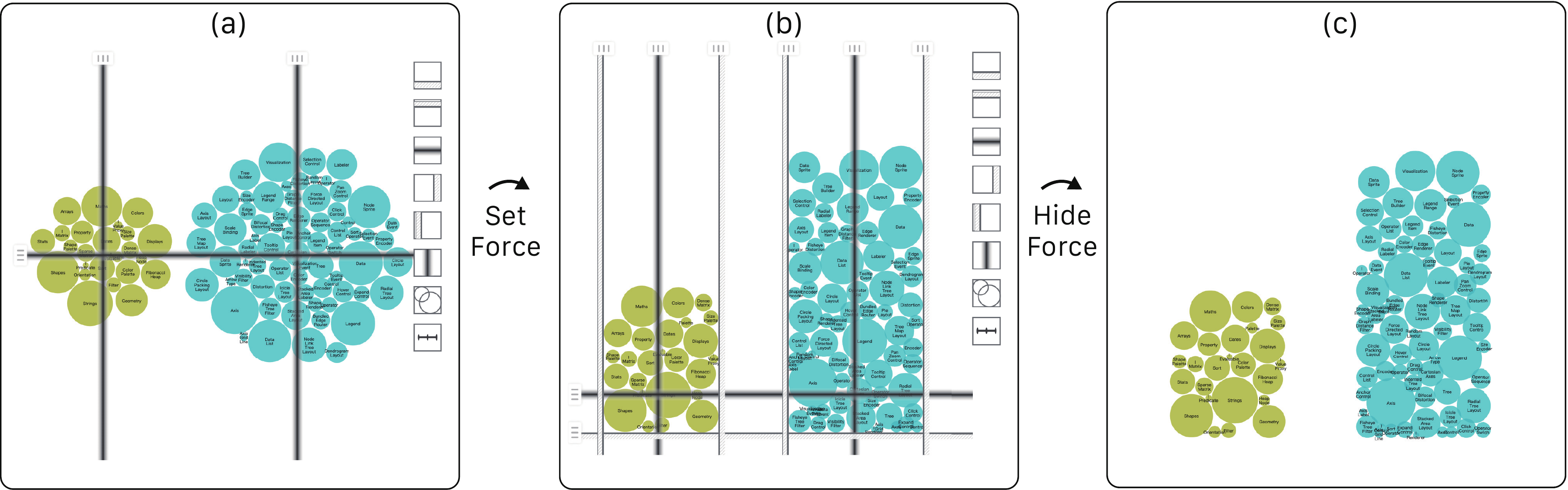}
    \caption{
        Manipulating constraints and setting new constraints for visual objects results in diverse new layouts of visual objects.
        (a) A bubble chart.
        (c) Constructing an aggregated view by setting support and gravity constraints.
    }
    \label{fig:force_handle}
\end{figure}

\subsection{Manipulating Constraints}
\label{sec:manipulate_cons}

The manipulation of constraints includes two parts: changing the existing constraints and setting new constraints.

\textbf{Changing constraints.}
The selected visual objects' constraints can be modified directly through dragging.
Our interface presents the top-$k$ gravity and support constraints with the most control points.
Each constraint has a handle for the users to manipulate.
When the constraints are dragged, related visual objects' constraints are changed.
For example, we can transform a stacked area chart into a ThemeRiver~\cite{havre2002themeriver} by dragging all visual objects' gravity constraints to the center, as illustrated in \autoref{fig:manipulate_classification} (f).
In \autoref{fig:manipulate_classification} (g), the bubbles can be squeezed by changing the support constraints.

\textbf{Setting new constraints.} 
Except for manipulating an existing constraint, users can set new constraints.
For instance, \autoref{fig:manipulate_classification} (g) and (f) can be interpreted as the establishment of new constraints.
Moreover, setting constraints on visual objects can construct diverse visualization layouts.
For example, as shown in~\autoref{fig:force_handle}, the bubble chart is reshaped to a bar chart by setting new support and gravity constraints.

\textbf{Setting groups of constraints.} 
When users want to rearrange visual objects based on their visual attribute values, they can assign different constraints to visual objects with different attribute values.
Considering that it is time-consuming to individually set constraints for each visual element, we allow users to batch set a group of different constraints based on a specific attribute (e.g., height, width, color).
The outcome is the segregation of visual objects with different attributes.
As depicted in \autoref{fig:manipulate_classification} (f), different $x$-axis gravity constraints can be assigned to bubbles of different colors.

\subsection{High-Level Interactions}

In subections~\ref{sec:manipulate_object}, \ref{sec:manipulate_axes}, and \ref{sec:manipulate_cons}, we describe low-level manipulations supported by our model.
These low-level manipulations are basic interactions for change the spatial layouts of the visualization.
These manipulations can compose high-level interactions for users' various requirements.
Users can perform various interactions using these manipulations.
We list interactions and discuss how they are composed by these manipulations.

\begin{itemize}

\item \textbf{Navigating} changes the users' viewpoints.
Navigating effectively narrows the field of view to allow users to observe details, e.g., in a dense scatterplot or a multi-line chart.
Visualization with continuous axes allows navigation interactions.
In our model, the navigation is performed by rescaling the continuous axes.

\item \textbf{Filtering} reduces the number of visual objects. 
Our method supports filtering by selecting focused visual objects and dragging them to a new canvas or by selecting unfocused visual objects and deleting them.
Filtering is a generic interaction for common visualizations such as, for example, bar charts, area charts, line charts, and scatterplots.

\item \textbf{Rearranging} changes the spatial organization of visual objects; it includes reordering, realignment, etc. 
Rearranging is supported by our model on three levels: there is object-level, axis-level, and constraint-level rearranging.
At the object level, users can drag visual objects to reorder, align, and stack them.
For example, users can reorder the categories of a ThemeRiver graph. 
At the axis level, manipulating discrete axes means reordering or sorting the axes.
At the constraint level, flexible constraint settings create a large space for rearrangement.
For example, a user can set support constraints for visual objects to align them.

\item \textbf{Re-encoding} represents to change the encoding of the visual objects.
Our model enables users to change encoding of visualizations at object level or constraint level.
At the object level, users can transform a grouped bar chart into a stacked bar chart or a stacked area chart into an overlapping area chart by simply drag on the visual objects.
At the constraint level, users can set groups of gravity constraints for visual objects according to their size or color, and they can re-encode their positions, as \autoref{fig:manipulate_classification} (h) shows, the constraint-level manipulation supports high freedom for users to define a new layout for the visualization.

\item \textbf{Aggregating} changes the granularity of the visual objects by gathering visual objects of the same type.
Users can set different gravity and support constraints at the constraint level for visual objects with different attributes, such as those with different colors. 
There is ample flexibility in setting constraints to aggregate certain visual objects.
For example, as illustrated in \autoref{fig:force_handle}, by setting collision, support, and gravity constraints, a bubble chart can be transformed into an aggregated bar chart.

\end{itemize}

\begin{figure}[!ht]
    \centering
    \setlength{\belowcaptionskip}{-10px}
    \includegraphics[width=1.04\columnwidth]{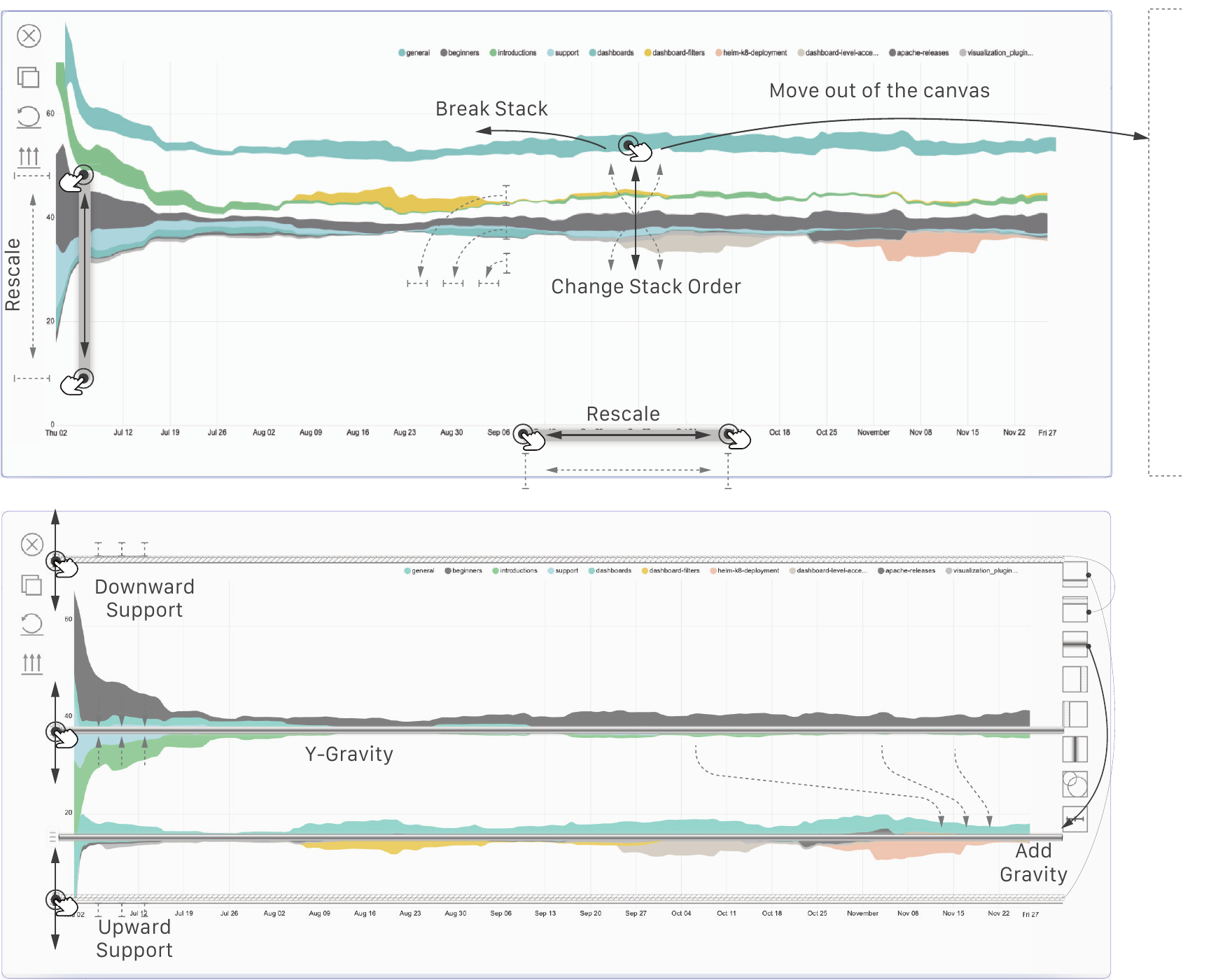}
    \caption{
        The interface of our system.
        Top: direct interactions with visual objects or axis.
        Bottom: direct setting constraints for visual objects directly.
    }
    \label{fig:interface}
\end{figure}

\subsection{Direct Manipulation Interface}
\label{section:interface}

The frontend interface is developed using JavaScript and the D3 library.
As illustrated in \autoref{fig:interface}, a visualization is rendered on a canvas.
To initiate the process, we create a canvas and reproduce the existing visualization on it.
Since we are redrawing the original visual objects using new paths, the corresponding visual objects are removed from the original visualization.
A new layer is introduced for visual objects and axes, employing control points to delineate these visual objects.
For area-type visual objects, we generate closed regions using the control points, while for line-type visual objects, we connect the control points.
Point-type elements are depicted as circular visual elements created using paths.
Visual attributes such as color, opacity, stroke color, and stroke width are preserved for these visual objects.
The resulting paths incorporate interactive features, enabling user actions like dragging.
Similarly, axes are fully redrawn using the D3 library, while maintaining the original axes' font styles and sizes.
The interface provides users with the ability to manipulate visual objects, axes, and constraints.

Furthermore, canvas operations have been implemented, including drag functionality, canvas creation, canvas duplication, canvas deletion, and scaling.
Four icons located in the top-left corner allow users to perform these actions, respectively: canvas deletion, canvas duplication, canvas reset, and display of the constraints layer.
Concerning the constraints layer, appropriate representations have been selected based on metaphors associated with different types of constraints.
Clicking the constraints button unveils a new layer that exhibits constraints and provides buttons for setting new constraints, as depicted at the bottom of \autoref{fig:interface}.
Constraints pertaining to the selected visual object are displayed and can be directly manipulated.
New constraints can be defined for the selected visual objects, encompassing support (upward, downward, left, rightward), gravity (vertical and horizontal), and collision constraints.

\section{Usage Scenarios} 
\label{section:use_scenario}

This section describes two real-world cases, including a stacked area chart from an online website and a bubble chart on a news website.

\begin{figure*}[!htb]
    \centering
    \includegraphics[width=\textwidth]{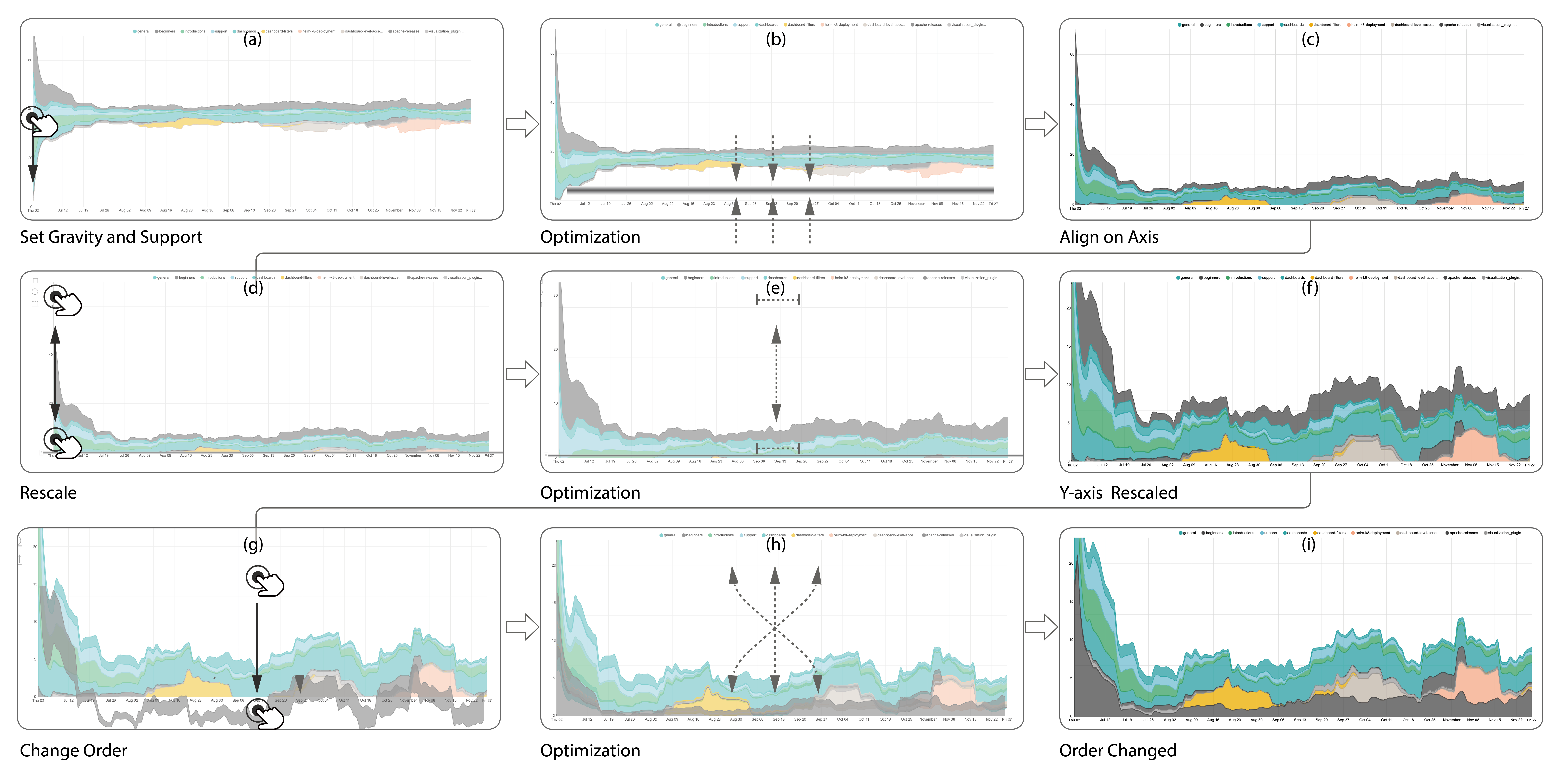}
    \caption{Stacked area chart example from an online webpage. 
    Our approach supports (a-c) aligning, (d-f)rescaling, and (g-i) reordering.}
    \label{fig:case_stack_chart}
\end{figure*}

\subsection{Stacked Graph of Slack Software's Message Trends}

While a stacked graph is a widely utilized visualization technique for illustrating temporal changes across multiple categories, it poses challenges related to legibility, comparison, and scalability~\cite{Baur2012touchwave}.
In terms of legibility, the perceptual distortion called the sine illusion effect~\cite{vanderplas2015signs, day1991sine} significantly affects the perception of values.
More specifically, within the context of a stacked area chart, distinct segments positioned atop it often lead to substantial perceptual distortions due to variations in the underlying stacking slopes.
Various algorithms~\cite{byron2008stacked, sinestream2021} have been developed to compute optimized static layouts for stacked graphs, aiming to enhance their legibility.
Furthermore, scalability becomes a concern when numerous categories are present, particularly when small values are challenging to discern.
Moreover, as a stacked graph represents aggregate values through the stacking of visual elements, comparing values across different time points or visual objects becomes intricate for users.
Our proposed approach addresses these aforementioned challenges by providing an effective solution for static stacked graphs.

We extracted the stacked graph that describes the trends of the messages per channel in the Slack software, as shown in \autoref{fig:case_stack_chart} (a); this graph comes from the Preset website\footnote{https://preset.io/blog/2020-09-22-slack-dashboard/}.
The original chart is a ThemeRiver~\cite{havre2002themeriver} without alignment on the $x$-axis.
A user, Martin, wants to uncover insights given this chart.
Our method makes it possible to improve the legibility of the stacked area chart by allowing the flex alignment of the visual objects and allowing the user to change the order of the visual objects.
We set an aligned baseline on the $x$-axis that is composed of two parts, the upwards support constraints of the $x$-axis and the gravity constraints of the $x$-axis.
After setting such new constraints, the visual objects fall, driven by gravity, as shown in \autoref{fig:case_stack_chart} (b), but they stop at the $x$-axis because of the support constraints from the $x$-axis.
The visual objects finally align and stack on the $x$-axis, as shown in \autoref{fig:case_stack_chart} (c), which results in a rearrangement that allows better value retrieval and the recognition of the trend for the total value.

In \autoref{fig:case_stack_chart} (c), the values after June 12th are relatively small compared to the beginning value.
These values only take up a small proportion of the vertical space, which makes it difficult to determine the data values.
Our approach allows Martin to rescale the chart in the y-direction to explore the visualization better, as shown in \autoref{fig:case_stack_chart} (d), (e), and (f).
The rescaling changes the height of the visual objects.
The ticks in the $y$-axis are updated according to the new scale, which is calculated using the D3 scale function.
However, some problems still occur because of stacking.
One problem is the distortion of the stacking area.
The mark on the top is heavily distorted because of different slopes.
As shown in \autoref{fig:case_stack_chart} (f), it is difficult for Martin to determine the trend of the top visual object (deep grey) because of the unalignment.
Martin can directly drag the visual object to the bottom to solve this problem.
The collision order of the visual objects is updated.
Collision constraints and support constraints cause the visual objects to converge at (i).
Consequently, the legibility, comparison, and scalability problems can be handled through these interactions.

\subsection{New York Times Vaccination Rate Bubble Chart}

Bubble charts are extensively discussed in academic papers (e.g., FluxFlow~\cite{fluxflow}) and are also widely used in news media.
Each point in the chart represents a data item.
Rearranging these data items can support various analytical tasks performed by users.

\begin{figure}[!ht]
    \centering
    \includegraphics[width=\columnwidth]{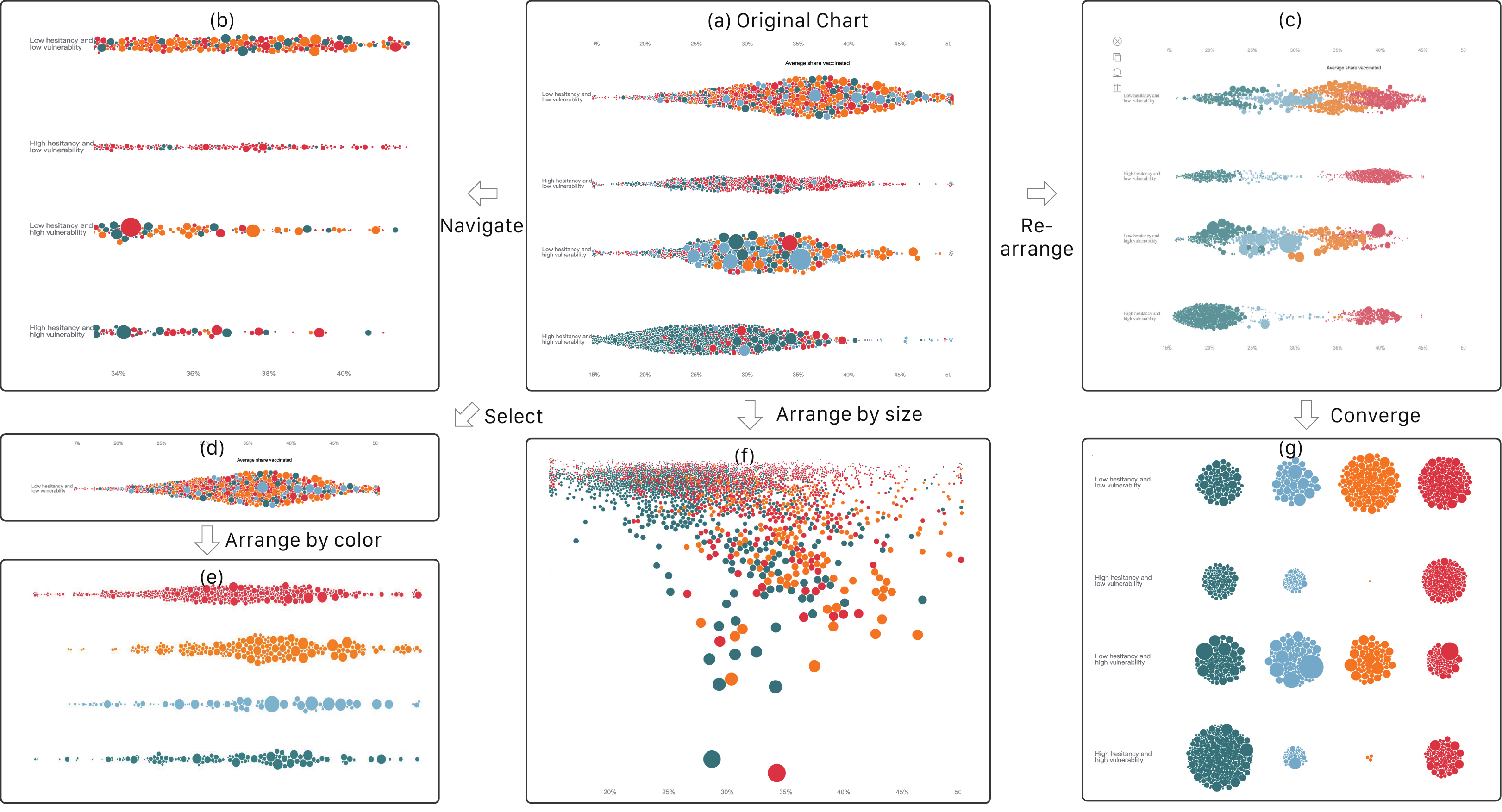}
    \caption{A bubble chart showing the vaccination rate of counties in the US. 
    Our method adds various interactions to the (a) original chart, supporting a wide range of tasks that involve manipulating the marks in the static visualization. 
    For example, in (a), (c), and (g), the visual objects are driven by different gravity constraints in the x-direction.}
    \label{fig:teaser}
\end{figure}

One example is the COVID-19 vaccination rate, which has received significant media attention.
The New York Times\footnote{\url{https://www.nytimes.com/2021/05/12/us/covid-vaccines-vulnerable.html}} published a chart that shows vaccine hesitancy, social vulnerability, and vaccination rates for each county in the US (\autoref{fig:teaser} (a)).
Each circle represents a county, with four different colors representing different regions (yellow: Northeast, blue: West, green: South, red: Midwest) and with the sizes of the circles indicating the population of each county.
In the original chart (\autoref{fig:teaser} (a)), from top to bottom, the counties are classified into four categories: ``low hesitancy and low vulnerability,'' ``high hesitancy and low vulnerability,'' ``low hesitancy and high vulnerability,'' and ``high hesitancy and high vulnerability.''

The chart encodes five data attributes: hesitancy, vulnerability, vaccination rate, population, and region.
Assume a user, Martin, reads the news on this website, and he wants to uncover insights by exploring the visualization.
Our approach can perform various manipulations of the chart to support data exploration.
Martin wants to explore the bubbles in detail. 
He can navigate the chart by zooming in on the $x$-axis, resulting in the view shown in \autoref{fig:teaser} (b).
In \autoref{fig:teaser} (b), he can see the details of specific counties' vaccination rates.
He wants to compare counties from different regions (colors), so he sets new x-gravity constraints in the x-direction that drives different colors to move to different positions.
The optimization process is shown in \autoref{fig:teaser} (c), where the points are moving to a new position.
The converged circles in \autoref{fig:teaser} (g) are rearranged by color; this is a kind of aggregation grouped by regions. 
Circles of the same region and category are aggregated together to estimate the population of each subset using the whole-area size of the group.
In the Northeast and West, most counties are in the low-hesitancy category, while in the South and Midwest, a large number of counties are in the high-hesitancy category.

Martin hypothesizes that counties with larger populations may exhibit higher vaccination rates. As a result, he intends to investigate the correlation between population size and vaccination rate. To achieve this, he establishes distinct y-gravity constraints corresponding to the population (radius). The resulting chart is depicted in \autoref{fig:teaser} (f), revealing a relatively weak correlation between population and vaccination rate. Nevertheless, counties characterized by low vaccination rates tend to be smaller in size. Since Martin resides in a county with limited hesitancy and vulnerability, he expresses a keen interest in the distribution of counties within the same category. He can effortlessly select this category by dragging it onto a new canvas, as showcased in \autoref{fig:teaser} (d). Additionally, he can configure y-gravity constraints for various colors (regions) as shown in \autoref{fig:teaser} (e).

These explorations yield Martin a deeper level of insight compared to the original chart. In conclusion, our approach empowers users to select, filter, arrange, and aggregate marks while retaining the contextual information present in the original chart.
Our model can also be directly applied to many other bubble charts. For instance, the bubble chart in the FluxFlow~\cite{fluxflow} can serve as an input for our model, allowing for rearrangement and recombination. Additionally, changes in Visual Sedimentation~\cite{Huron2013Visual} that do not involve changes in the type of visual elements (such as from bubbles to area sedimentation) can also be represented using various constraints.

 \section{User Interview}
\label{section:user_study}

We conduct a user interview to evaluate whether or not our approach provides users with effective interactions and to see how our methods can support users' analytical tasks.
As no previous work presented such interaction enhancement, we did not compare our approach with other studies.
From the creator's perspective, our method eliminates the need for writing interactive code, while from the user's standpoint, it is easy to comprehend. Therefore, our model is suitable for users with varying levels of experience. Consequently, we have invited participants with different levels of experience to engage with our method.

\subsection{Study Design}

\textbf{Participants}. 
We conducted a user study with a cohort of 10 participants (P1-P10).
These participants possessed varied educational backgrounds, spanning fields such as information science, data science, computer graphics, mathematics, among others.
We collected self-reported measures of their experience in the realm of visualization, assessed using a 5-point Likert scale.
This encompassed aspects such as their comprehension of data visualization (midian = 4, range = 2, IQR = 1),
familiarity with software-based visualization tools (midian = 3.5, range = 2, IQR = 1),
as well as utilization of programming for generating visualizations (midian = 3, range = 3, IQR = 0.75).

\textbf{Training and tasks.}
Prior to the commencement of the study, participants were furnished with a concise overview and demonstration of the system. Essential functionalities were expounded, encompassing the manipulation of visual entities, coordinate axes, and imposed constraints.
Subsequently, participants were allotted a span of 10 to 20 minutes to organically explore the system, acquainting themselves with its diverse capabilities. Following this preliminary training phase, participants were introduced to functionalities that had evaded their initial exploration.
Participants were tasked with manipulating two visualizations: the stacked area chart and the bubble chart elucidated in Section~\ref{section:use_scenario}.
Employing our system's manipulation features, participants were tasked with devising novel visual layouts and employing these configurations to fulfill specific tasks.
These tasks encompassed customary visualization undertakings as delineated in the literature~\cite{Brehmer2013TaskAbstract}, inclusive of identification, comparison, and summarization.
For the bubble chart, four tasks were presented:
\begin{itemize}
    \item \textbf{T1:} Present the ranking of different regions with vaccination rates between 30\% and 35\%.
    \item \textbf{T2:} Identify which region has the most counties with low vaccination rates.
    \item \textbf{T3:} Identify whether a county's vaccination rate is correlated to its population.
    \item \textbf{T4:} Identify which region within the high hesitancy and low vulnerability category has the largest population.
\end{itemize}
For the stacked area chart, we provided four tasks:
\begin{itemize}
    \item \textbf{T5:} Identify the maximum among the categories on June 12th.
    \item \textbf{T6:} Compare the values of the yellow category (i.e., dashboard filters) on Aug 16th and Aug 30th.
    \item \textbf{T7:} Present the trend of the combined values within the deep grey and yellow categories.
    \item \textbf{T8:} Show the difference in the trends of the deep grey and yellow categories.
\end{itemize}

\begin{figure}[htb]
    \centering
    \includegraphics[width=\columnwidth]{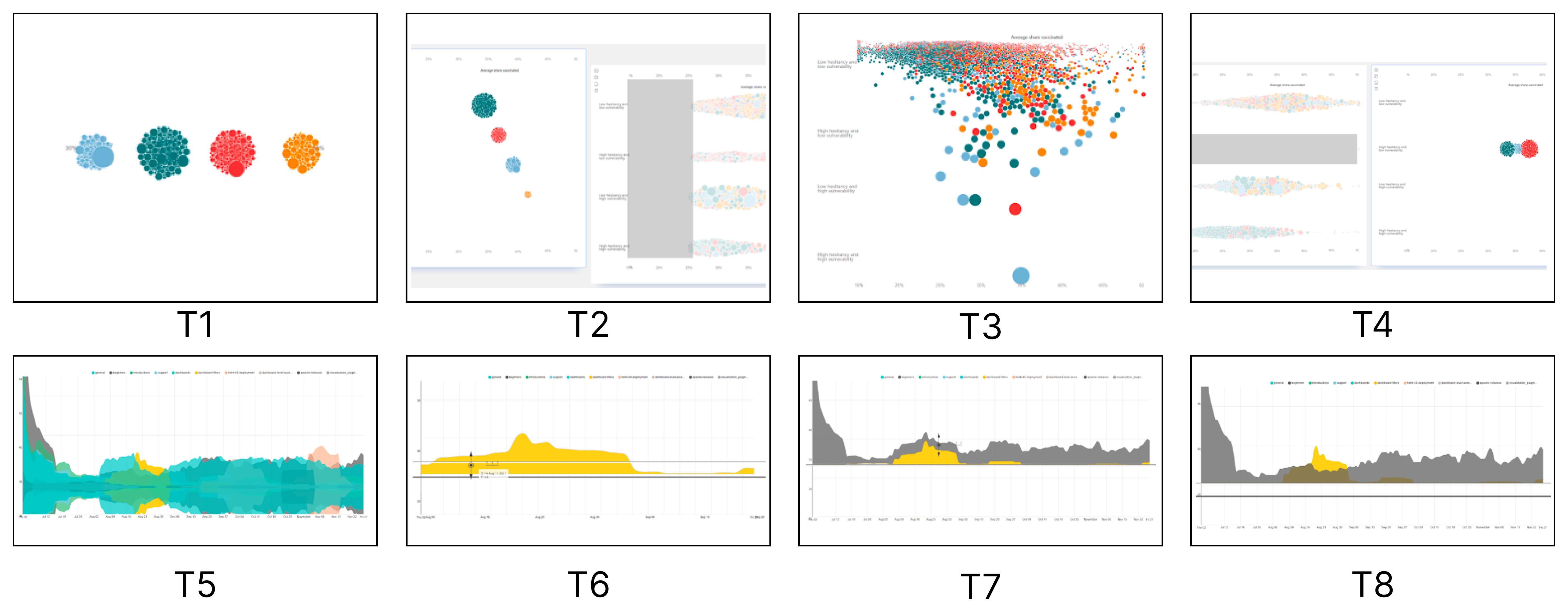}
    \caption{
    Representative results for different tasks.}
    \label{fig:results}
\end{figure}

\begin{figure}[htb]
    \centering
    \includegraphics[width=\columnwidth]{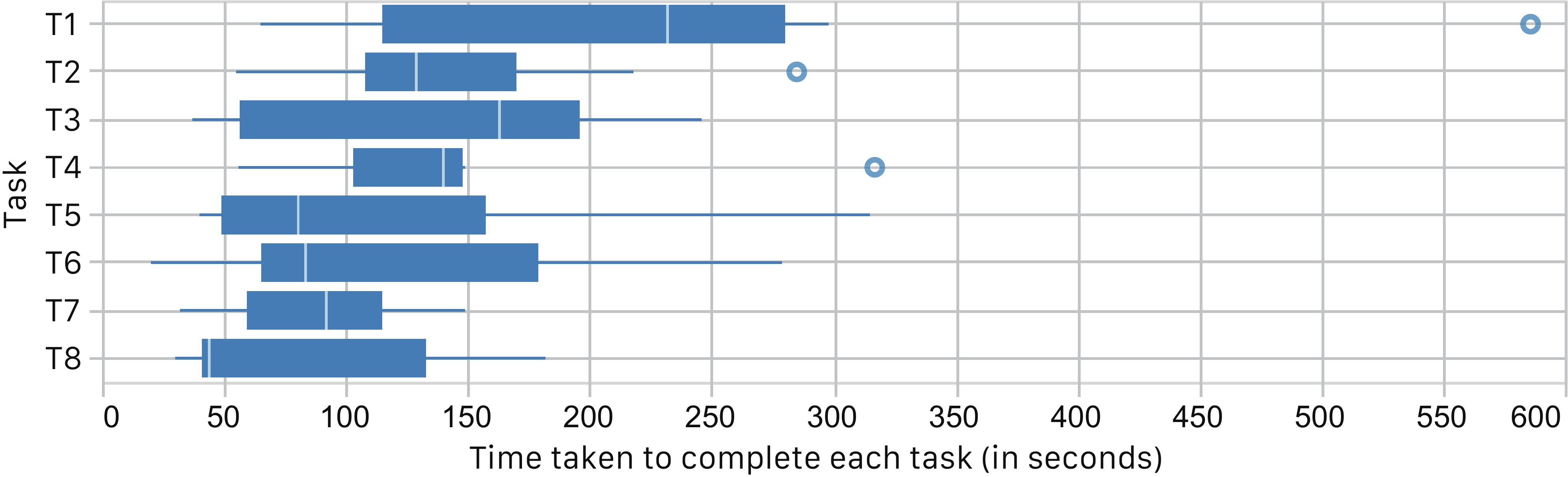}
    \caption{
    The time taken for different tasks. It is evident that tasks T1-T4 require a longer time to complete compared to tasks T5-T8.
    }
    \label{fig:costtime}
\end{figure}

\begin{figure}[htb]
    \centering
    \includegraphics[width=\columnwidth]{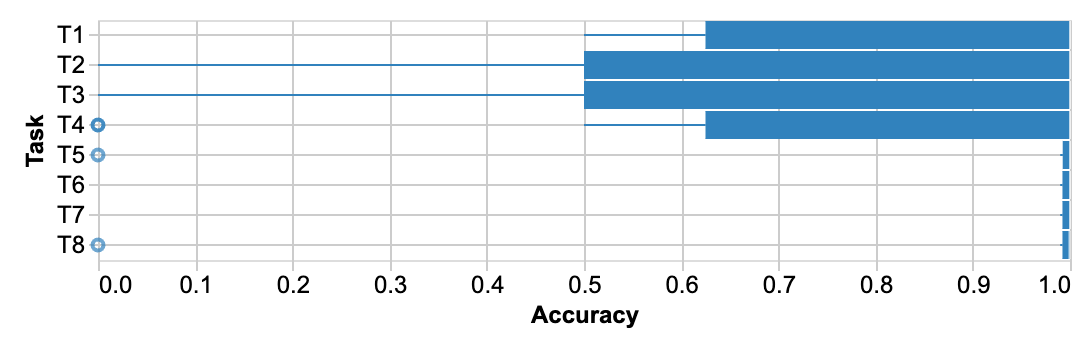}
    \caption{
    The completion rate for different tasks. Tasks T1-T4 have lower completion rate compared to tasks T5-T8.
    }
    \label{fig:completion}
\end{figure}

Upon completing the tasks, we requested participants to upload screenshots of their manipulated visualizations and elucidate the rationale behind their chosen layout arrangements.
\autoref{fig:results} shows the representative results for 8 tasks.
On average, participants took approximately 20 minutes to conclude their user study.
We gauged the efficacy of user-uploaded screenshots in facilitating task comprehension.
For example, in Task 8, we anticipated participants to align and compare content from two distinct categories. Ratings were as follows: 1 point for complete task alignment, 0 points for lack of task alignment, and 0.5 points for partial task resolution.
Ultimately, the majority of participants successfully completed most tasks, yielding an average score of 85\%. User scores ranged from a high of 100\% (P2) to a low of 62.5\% (P5). Among the eight tasks, the average score for the four bubble chart tasks was 75\%, while the four stacked area chart tasks averaged 95\%. The stacked area chart significantly outperformed the bubble chart in performance.
This observation finds further support in the time spent and completion rate, as portrayed in \autoref{fig:costtime} and \autoref{fig:completion}. 
The time dedicated to bubble chart tasks (T1-T4) notably exceeded that of stacked area chart tasks (T5-T8) by a substantial margin.
This disparity is attributed to the fact that bubble chart tasks typically entail constraint establishment, whereas tasks linked to the stacked area chart commonly involve drag-and-drop interactions.
Subsequently, we administered a subjective-rating questionnaire and conducted interviews to gather feedback on the system's efficacy.

\subsection{Feedback from Participants}

We solicited participant feedback regarding the comprehensibility of the optimization process after manipulation, the extent to which the manipulated layout facilitated user tasks, and the comprehensibility of representing visual objects with spatial constraints. Each aspect was evaluated using a 5-point scale (1: not effective at all, 2: not very effective, 3: neutral, 4: somewhat effective, and 5: very effective). The results demonstrated favorable responses across multiple dimensions.
The user feedback is summarized as follows:

\textbf{Manipulations are easy to learn and effective.}
Participants commonly employ our method to manipulate visualizations for task completion. Furthermore, users generally perceive that the manipulated visual outcomes aid in task accomplishment ($\mu=4.0$, $\sigma=0.82$).
Participants used a range of manipulation techniques, such as dragging selected data items to a new canvas (T1, T4, T7, T8), rearranging vertical or horizontal axes based on the color or size channel (T1, T2, T3, T4), breaking stacking relationships to facilitate comparisons (T5 and T8), and setting constraints to align visual objects (T6).
Participants emphasized that the system to be comprehensible (P2, P7). P2 suggests that \textit{``the system is highly beneficial for users to understand visualizations, as users can leverage their understanding of elements from the physical world to comprehend visualization elements, thereby facilitating a quicker grasp of visualization interactions.''}

\textbf{The position update process is intuitive.}
Participants demonstrated a high level of comprehension regarding the optimization process ($\mu=4.30$, $\sigma = 0.67$).
Several participants highlighted that the procedure for updating the positions of visual objects exhibited an intuitive nature and remained in line with their anticipations (P2, P4, P6, P7).
P7 particularly underscored that \textit{``the operations align with common-sense notions of physical principles, making the process of change easily understandable.''}
The clarity of the optimization process was evident in participants' grasp of the physical forces employed to simulate the interactions, a factor that significantly contributed to their comprehension of the positional adjustments within the visualization.

\textbf{Manipulating constraints increases flexibility.}
Many participants perceive the system as addressing the flexibility of the visualization interaction space (P1, P2, P5, P8, P9). For instance, P9 mentions that ``with the system, tasks can be accomplished more freely without relying on the original visualization interaction. The layout can be designed freely.'' These flexible approaches aid task completion (P5, P10). However, some users also note that while utilizing constraints significantly enriches the interactive operational space, facilitating more flexible adjustments to data distribution, it might pose challenges for newcomers. They express the desire for additional interactive cues that encompass the potential outcomes following operations and the interactive tasks that could be supported.

 \section{Discussion and Future Work}
\label{section:discussions}

In this section, we discuss the coverage of the supported interactions and supported visualization and point out directions for future work.

\subsection{Coverage of Supported Interactions}

The manipulations supported for an existing visualization are contingent on the current set of visual objects within that visualization.
There exist two constraints: our methodology does not alter the quantity of control points, and it solely relies on information present in the current visualization without any additional data.
Of the unipolar interactions as categorized by Sedig and Parsons~\cite{sedig2013interaction}, operations such as drilling and blending, which necessitate supplementary data or a change in the control point count, are not compatible with our framework. Conversely, arrangements, assignments, cloning, comparisons, filtering, navigation, transformations, and translations can be partially or fully facilitated within the present framework, as these actions can be translated into alterations in the positions of control points.
In the realm of bipolar interactions, composing and decomposing provoke a change in the number of control points, making them non-applicable within our framework.
Interactions like gathering and discarding, inserting and removing, as well as storing and retrieving, can be executed by manipulating constraints and relocating visual objects between canvases.
In the future, the extensibility of the constraints model could involve permitting changes in control point quantities and associating visual objects with additional information.

\subsection{Coverage of Supported Visualizations}

In the previous sections, we demonstrated some common visualizations, including area charts and bubble charts.
Our approach can be applied to more visualizations.

\textbf{Limitations arising from implementation.}
Currently, our focus in terms of implementation is on non-nested visualizations within the two-dimensional Cartesian coordinate system.
The current implementation of this study effectively encompasses various prevalent visualizations, including scatter plots, line charts, and bar charts (both simple and with stacked or grouped variations).

\begin{itemize}
\item \textbf{2D Cartesian coordinates:}
The prototype system is primarily focused on visualizations within the two-dimensional Cartesian coordinate system.
These visualizations involve visual elements distributed either discretely along axes at fixed frequencies, as seen in heatmaps or Bertin Matrices~\cite{perin2014revisiting}, or more irregularly across continuous axes. We start by extracting both the horizontal and vertical axes. Then, we analyze the distribution patterns of visual elements along these axes.
By integrating these patterns with the axis structures, we are able to derive constraints for the visualizations.

\item \textbf{Exclusion of nested visualizations:}
The current model implementation does not encompass multi-layer nested structures. It does not yet accommodate visualizations that involve glyphs or exhibit nested properties, such as a bar within a grouped bar chart composed of multiple stacked bars. However, this model can accommodate such visualizations in future extensions.

\item \textbf{Deviation from conventional rules:}
Instances of parsing errors predominantly stem from heuristic solving algorithms. These errors might involve misinterpretation of numerical axes or visual arrangement patterns that deviate from common conventions.
\end{itemize}

Potential future extensions may encompass polar coordinate systems. In this context, sector charts could be interpreted as bar charts, and donut charts as stacked bar charts. Such extensions would require the augmentation of parsing algorithms for polar coordinates, as well as the development of coordinate system conversion algorithms.

\begin{figure}[htb]
    \centering
    \includegraphics[width=\columnwidth]{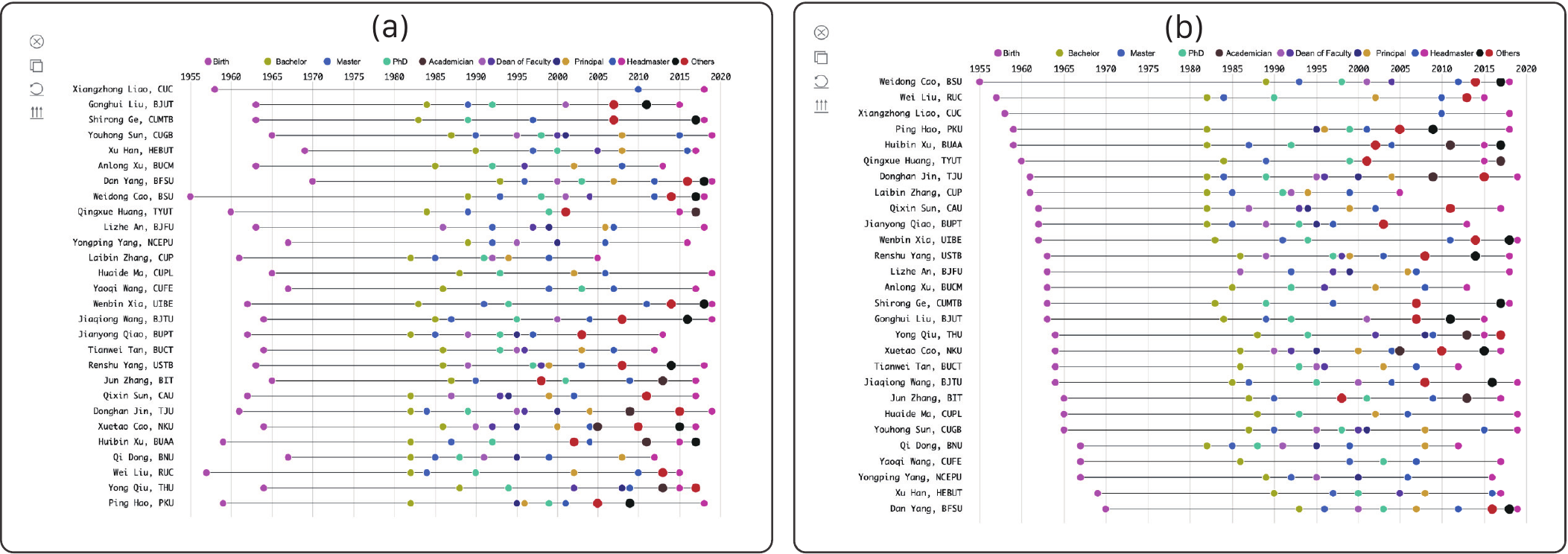}
    \caption{(a) The original static visualization presents important moments in the lives of university presidents.
    (b) The view after they are sorted according to their birthdays.
    }
    \label{fig:class_project}
\end{figure}

\textbf{Coverage of visualization types.}
Ideally, at the conceptual level, common visualizations encode spatial channels with attributes that are compatible with our spatial constraints model.
For example, the interactions that change the stacked order/direction will be effective for the visualizations with stacked collisions.
For those visualizations that only have gravity constraints, including line charts and scatter plots, selecting and filtering can be supported by moving visual objects to a new canvas.
Currently, visualizations with a Cartesian coordinate system can be parsed by our system.
The axis type determines what axis operations are supported.
For example, the visualization in \autoref{fig:class_project} shows important moments in the lives of China's university presidents (e.g., being born or obtaining a Ph.D. degree).
We can sort the presidents according to their birthdays to better support comparison tasks.
The results are shown in \autoref{fig:class_project} (b).
We can extract a horizontal axis and a vertical axis (i.e., a visualization with a Cartesian coordinate system), and the continuous axis is taken as the linear scale. 
In the future, we could extend the model to visualizations with parallel axes or polar coordinate systems. 
For parallel axes, spatial constraints can be constructed for every axis.
If we want to support manipulations that change the organization of the axes (e.g., change the axis order of parallel coordinates), additional rules for axis positioning are needed.
For polar coordinate systems, we can conceptually regard the coordinate systems as Cartesian coordinate systems in which the pie chart is stacked in the angle direction.

\textbf{Raster image support.}
Our current prototype system primarily focuses on parsing visualizations in vector formats. However, many static visualizations are not represented in vector formats, such as historical visualizations~\cite{zhang2023oldvis}. For such visualizations, the automatic addition of interactivity would be of great interest. In the future, we envision leveraging object detection and optical character recognition (OCR) techniques to support raster images. Subsequent parsing procedures, akin to the present modeling process, will ensue after control points extraction. This approach holds the potential to significantly broaden the scope of applicable scenarios.

\subsection{Toward interaction authoring toolkit}

The primary innovation of our work lies in enhancing the interactive capabilities of existing visualizations. This approach emphasizes spatial representation within visualizations while disregarding tools used for visualization creation.
The principal challenge in extending this work into a toolkit lies in compatibility. The fundamental unit of interaction and transformation in this paper is the control point. However, if one aims to integrate corresponding plugins into existing toolkits, a reconfiguration of the original visualization implementation in terms of control points is required. Such an implementation may encounter resistance from existing visualization practices; for instance, within visualizations realized using Vega-Lite, our approach necessitates a reorganization of visualizations.

Customized interactive capabilities will be provided, catering to the specific needs of scenarios. The current method offers a comprehensive range of interaction features for existing visualizations. It supports the addition of interactive features to these visualizations. For many toolkit scenarios, the data and toolkit used in visualizations are known. As a manifestation of a toolkit, this paper is a well-suited candidate for augmenting interactivity upon pre-existing visualizations, with provisions for user customization, such as defining required interactions or enabling user-defined settings.

Our model enables the activation of static visualizations through a set of low-level operations involving visual objects, coordinate axes, and constraints. Nevertheless, these low-level interactions might pose a time-consuming obstacle to users in constructing meaningful interactions. To address this issue, for instance, a set of constraints could be established to transform bubbles of different colors into bars within a bar chart, obviating the need for individual constraint configuration, as illustrated in Figure \ref{fig:force_handle}. In the future, incorporating user-defined constraints as novel forms of interaction and preserving them could significantly augment the semantic diversity of interactions.

\subsection{Toward Intelligent Interaction}

This method can be enhanced in several aspects through the application of deep learning techniques.

\textbf{Intelligent constraint deduction.}
During the constraint inference process, it is common to employ heuristic and rule-based algorithms to cover prevalent implementation methodologies. However, due to varying user implementations, inherent uncertainty emerges in the inference process. For instance, within a ThemeRiver visualization, the mapping of data based on relative height versus absolute height may lack certainty. The incorporation of deep learning models can aid in resolving such classification ambiguities.

\textbf{User intent inference through mixed initiatives.}
Concerning the aspect of user intent inference, while achieving complete machine-driven inference is challenging, a subset of classification tasks can be formulated to guide machine decision-making. For instance, one such task involves discerning whether a user rapidly drags a visual element beyond the current canvas – an action open to diverse interpretations of speed across user profiles. To address this, user feedback can be leveraged, enabling deep learning to enhance user exploration efficiency. It could contribute to decisions like creating new canvases, deleting visual objects, among other actions, while not entirely replacing user agency.

\textbf{Combining our model with other interaction techniques.}
Our method facilitates a more intuitive interaction for users to comprehend visualizations. This endeavors to minimize the user's barrier to utilization and comprehension of visualization interactions. In the future, incorporating natural language interaction represents another means of reducing the user's threshold. Our approach can be integrated with natural language interaction techniques, enabling users to simply describe the comparison they wish to make between two parts of a stacked area chart, for example. After natural language processing, our system can then further display the transition of visual objects that clearly illustrates the differences. This approach results in reduced barriers for the user in terms of both articulating their intention and comprehending changes.

 \section{Conclusion}
\label{section:conclusions}

We introduced a spatial constraint model to characterize the spatial positions and spatial changes of visual objects.
The spatial constraints are translated into forces that guide the convergence of control point positions as users manipulate visual objects, coordinate axes, and constraints.
Leveraging the spatial constraint model, we developed a prototype system that seamlessly integrates intuitive interactions into preexisting visualizations.
This system facilitates direct manipulation of visual objects, coordinate axes, and constraints, resulting in novel visualization layouts following convergence, thus accommodating diverse interaction tasks.
We demonstrated the capabilities of the model through multiple usage scenarios and a user study.
The findings indicate that the optimization process is intuitive and that the interactions yield effective layouts for various tasks.

\begin{acks}
This work is supported by NSFC No. 62272012. 
\end{acks}

\bibliographystyle{ACM-Reference-Format}
\bibliography{manuscript.bib}

% \printbibliography

\end{document}